\documentclass[12pt,onecolumn]{IEEEtran}

\pdfoutput=1

\usepackage{bm}
\usepackage{psfrag}
\usepackage[mathscr]{eucal}
\usepackage{times}
\usepackage{epsfig,graphicx,color,pstricks}
\usepackage{amsmath,amssymb,amsthm,amsfonts,eucal,latexsym}
\usepackage{cite}
\usepackage[all]{xy}
\usepackage{subfigure}
\usepackage{graphics}

\newcommand{\CC}{\mathbb{C}}
\newcommand{\RR}{\mathbb{R}}

\newcommand{\NN}{\mathbb{N}}

\newcommand{\diff}{\mathrm{d}}
\newcommand{\eu}  {\mathrm{e}}
\newcommand{\jj}  {\mathrm{j}}

\newcommand{\snr} {\mathsf{SNR}}

\renewcommand{\Pr} {\mathop\mathrm{Pr}\nolimits}

\newcommand{\E}  {\mathbb{E}}

\newcommand{\reff}[1]{\eqref{#1}}
\newcommand{\defeq}{\triangleq}

\newcommand{\Xv}{\boldsymbol{X}}
\newcommand{\Yv}{\boldsymbol{Y}}
\newcommand{\Zv}{\boldsymbol{Z}}

\newcommand{\Pc}{\mathcal{P}}

\newcommand{\Rc}{\mathcal{R}}

\newcommand{\Tc}{\mathcal{T}}

\newtheorem{theorem}{Theorem}

\newtheorem{proposition}[theorem]{Proposition}
\newtheorem{corollary}[theorem]{Corollary}

\begin{document}

\title{On the Benefits of Partial Channel State Information for Repetition Protocols in Block Fading Channels}

\author{Daniela Tuninetti\\
\thanks{Dr. D. Tuninetti is with the Electrical and Computer Engineering Department
of the University of Illinois at Chicago, Chicago, IL 60607, USA.
Parts of this work were presented at~\cite{tuninettiICC2008,tuninettiITW2007}.
This work was partially funded by NSF under award number 0643954.
The contents of this article are solely the responsibility of the authors and
do not necessarily represent the official views of the NSF.}
}

\maketitle

\begin{abstract}
This paper studies the throughput performance of
HARQ (hybrid automatic repeat request) protocols over block
fading Gaussian channels.  It proposes new protocols that use the
available feedback bit(s) not only to request a retransmission, 
but also to inform the transmitter about the instantaneous channel quality.
An explicit protocol construction is given for any number of
retransmissions and any number of feedback bits.
The novel protocol is shown to simultaneously realize
the gains of HARQ and of power control with partial CSI
(channel state information).
Remarkable throughput improvements are shown, especially at low 
and moderate SNR (signal to noise ratio),
with respect to protocols that use the 
feedback bits for retransmission request only.
In particular, for the case of a single retransmission and a single feedback bit,
it is shown that the repetition is not needed at low $\snr$  where the
throughput improvement is due to power control only.  
On the other hand, at high SNR, the repetition is useful and 
the performance gain comes form a combination of power control
and ability of make up for deep fades.
\end{abstract}


\section{Introduction}
\label{sect:into}

\IEEEPARstart{I}{n}  today networks, error correction is achieved by a combination of
FEC (forward error correction) and ARQ (automatic repetition request).
In classical ARQ protocols, a receiver requests a
retransmission (sends a negative acknowledgment, or NACK)
when an error is detected, and  a positive acknowledgment (ACK) otherwise.
In this work we explore the performance gain achievable by using the
retransmission request bit(s) to signal to the transmitter the
decoder status {\em and} the actual channel state, albeit coarsely.
Our goal is to simultaneously enable the performance
gain due to HARQ (hybrid automatic repeat request),
i.e., a combination of ARQ and FEC error control methods~\cite{caire_tuninetti:arq_it},
and to power control at the transmitter~\cite{goldsmith_varaiya}.

\subsection{A Motivating Example}

Consider a fixed rate transmission scheme over a block
fading Gaussian channel with unit noise power spectral density.
Let the transmit power in slot $t$, $t\in\NN$, be $Q_t$,
the fading power gain be $\gamma_t$,
and the transmission rate be $R$.
The receiver fails to decode when the instantaneous channel
capacity is below the transmission rate~\cite{summa_fading_bc},
in which case it feeds back a NACK to the transmitter.  A NACK is thus equivalent to
\[
\log(1+\gamma_t Q_t)< R
\Longleftrightarrow
\gamma_t <\frac{\eu^R-1}{Q_t},
\]
that is, a NACK is a {\em 1-bit quantization of the channel state information} (CSI)
$\gamma_t$ sent to the transmitter when the transmitter no longer needs
it (as already remarked in~\cite{Kim-Skolunt,elgamalcairedamen:dmtharq:it06}).
This simple observation raises the question investigated in this work: 
whether it is optimal, in some sense, to feedback a ACK/NACK at the end of a slot,
or whether the same feedback resources should rather be used at the beginning of the slot
to inform the transmitter about the instantaneous channel quality, albeit coarsely.

To further gain insights into the problem, consider the outage capacity~\cite{summa_fading_bc}
as the performance measure.  For a fixed positive parameter $s$, 
let the transmission rate be parametrized as $R=\log(1+\overline{P}\,s)$,
where $\overline{P}$ denotes the average $\snr$  $\overline{P}=\E[Q_t]$.
As explained before, a 1-bit feedback used for ACK/NACK at end of slot $t$
indicates to the transmitter that $\gamma_t<s$ if a NACK is received,
or that $\gamma_t\geq s$ if a ACK is received.  The probability of
successful decoding is then the probability of receiving a ACK;
thus the outage capacity, or long term average successfully decoded rate, is
\[
\eta_{\rm ACK}=\Pr[\gamma_t\geq s]\log(1+\overline{P}\,s).
\]
On the other hand, consider the case where
the 1-bit of feedback is used at the beginning of the slot
to indicate to the transmitter which of the events, $\{\gamma_t< s\}$
or $\{\gamma_t\geq s\}$, has occurred.
In this case the transmitter can use this information as follows.
It turns transmission off (i.e., $Q_t=0$) if the channel is bad (i.e., $\gamma_t< s$) and
it sends with power $Q_t = \frac{\eu^R-1}{s}$ if the channel is good (i.e, $\gamma_t\geq s$);
the chosen transmit power $Q_t= \frac{\eu^R-1}{s}$ is such that no outage occurs.
The average transmit power of this simple
power control policy based on 1-bit CSI
is $\overline{P} = \frac{\eu^R-1}{s}\Pr[\gamma_t\geq s]$ and
the outage capacity is
\[
\eta_{\rm CSI}=\Pr[\gamma_t\geq s]\log\left(1+\overline{P}\frac{s}{\Pr[\gamma_t\geq s]}\right).
\]
It is immediate to see that, for the same set of parameters
$\overline{P}$ and $s$, and with 1-bit of feedback in both scenarios,
the outage capacity with CSI $\eta_{\rm CSI}$ is larger than the outage capacity
with ACK/NACK $\eta_{\rm ACK}$.  
This observation reinforces the idea that using
the 1-bit feedback to signal ACK/NACK is not optimal
in general. 
The question whether this conclusion changes if retransmissions are
allowed is investigated in this paper.

\subsection{Past Work}
To the best of the author's knowledge, past work available in the literature
considering quantized and/or noisy CSI
only focused on outage capacity, or on outage probability,
or on expected capacity, but not on HARQ protocols.

For example, in~\cite{Bhashyam} the authors consider
power control policies for minimizing the outage probability
with partial CSI;  the derived power
policy shows benefits with respect to the case
of complete absence of CSI even if the
channel knowledge is noisy and/or partial;
the benefits are more pronounced at low $\snr$.
In~\cite{steinber-samai}, the authors considered the outage capacity
with the so called ``broadcast approach'', that is, a multiple-layer
coding scheme with infinite many layers, where the receiver decodes as many layers
as possible given the actual channel fading;
it is found that even 1-bit of CSI helps to improve performance.

In~\cite{caireshamai:capacitypartialcsi}, the authors studied
the ergodic capacity of channels with states where the state
is only partially known at the transmitter; for the Gaussian
channel with quantized CSI, they showed that the 
capacity achieving power allocation is of the waterfilling type.
In~\cite{Kim-Skolunt}, the authors studied the expected capacity
with quantized CSI and multiple-layer coding schemes;
they showed that multiple-layer transmission offers limited
benefits when power control at the transmitter is possible.

In~\cite{elgamalcairedamen:dmtharq:it06}, the authors considered the DMT (diversity
multiplexing tradeoff) of multi-antenna channels with HARQ; in this
setting the feedback is only used to signal ACK/NACK and not
to perform power control, even though the transmitter
is allowed to vary the transmit power across retransmissions;
it is found that HARQ improves
the DMT by a factor proportional to the maximum number of repetitions.

\subsection{Contributions}

In this work we consider the joint design of HARQ protocols
and power control for block fading Gaussian channels.
We use the {\em long-term average decoded rate}~\cite{caire_tuninetti:arq_it},
or simply {\em throughput} for brevity in the following, as a measure of performance.
The throughput captures the fundamental performance limits when
strict delay constraints are imposed and includes the outage capacity
and the ergodic capacity as special cases.

When considering HARQ protocols, it is customary to
assume that the transmitter has no knowledge of the instantaneous fading and
it thus transmits with equal power in every slot~\cite{caire_tuninetti:arq_it}. 
However, especially with INR (incremental redundancy), the probability of having to transmit
$m$ channel packets per data packet is decreasing with $m$~\cite{elgamalcairedamen:dmtharq:it06}.
Thus, it is conceivable that using more power in earlier transmissions
of the same data packet reduces the probability of decoding failure and hence 
increases the throughput for the same average transmit power.
Moreover, if the fading is known at the transmitter, more power
can be used in the most favorable channel conditions--assuming power control is possible.
As pointed out in~\cite{caire_tuninetti_verdu:rate_it}, in delay constrained scenarios,
the assumptions about the dynamics of the fading process with respect to the
code length, as well as the duration over which power constraints are enforced,
are critical.  Here, in order to enable power allocation,
we consider a power constraint imposed over
a time horizon comprising many slots (i.e., much bigger than the maximum number
of retransmissions allowed), commonly referred to as {\em long-term average power
constraint}~\cite{caire_taricco_biglieri:optimal_powercontrol}.

As opposed to classical HARQ protocols, where the ACK/NACK feedback bit
is sent at the end of the slot, we consider here systems where the feedback bit(s) can be sent back 
at any point in time during a slot; the feedback can be used to signal CSI, or ACK/NACK, or any combination of them. 
The only restriction we impose is that the feedback bit(s) cannot be carried over from slot to slot.

Our contributions can be summarized as follows:
\begin{enumerate}
\item
We propose novel HARQ protocols where the CSI and the ACK/NACK information are combined within the same feedback bit(s) in order to {\em realize simultaneously the gains due to HARQ and the gains of power control with partial CSI}. 
The main idea behind the proposed protocols is that the receiver sends back to the transmitter
the index of the smallest power level that will allow successful decoding in the current slot.
Our protocols are time-varying quantizers for a suitably
scaled version of the channel fading, where the scaling factor accounts for the information
already available at the receiver from the past transmissions.

\item
We show that the throughput performance of the proposed class of protocols with {\em perfect CSI}
can be obtained from dynamic programing~\cite{book:kumar_varaiya:stochastic_systems}.

\item
By numerical evaluations of the throughput for Rayleigh fading channels,
we show that repetitions are not needed at low $\snr$, and that the
improvement over classical HARQ protocols (that use the feedback bit for ACK/NACK only)
is entirely due to ability to perform power control. 
 
At high $\snr$, repetitions are useful, and 
the performance improvement over classical HARQ
protocols comes form a combination of power control
and the ability to make up for atypical long deep fades with repetitions.

Our numerical results show that our protocols outperform classical HARQ
at all $\snr$s.

\item
We also have the following side results: 
(a) we show that the optimal power
allocation for the outage capacity with partial CSI
consists of a quantizer of the fading gain where the
quantization regions are union of intervals, rather than intervals;
to the best of the author's knowledge--was not reported before;
(b) we present novel bounding techniques for to compute
certain probabilities that are needed for the
throughput evaluation; these techniques are useful
for numerical optimizations. In particular, a technique
based on considerations on the order statistics of
an independent sample of negative exponential random variables
is of interest in its own.

\end{enumerate}

\subsection{Paper Organization}
The rest of paper is organized as follows.
Section~\ref{sect:model} introduces the system model and 
Section~\ref{sect:eta eval} evaluates the throughput;
Section~\ref{sec:eta M=infty,F,(INR)} revises the ergodic capacity with partial CSI,
which serves as an upper bound for any HARQ protocol;
Section~\ref{sec:eta M=1,F,(ALO)} derives the outage capacity with partial CSI,
which serves as a lower bound for any HARQ protocol;
Section~\ref{sect:main} proposes a new class of HARQ protocols that
combine repetitions and power control for any number of retransmissions
and any number of feedback bits;
Section~\ref{sect:main bounds} proposes a novel 
bounding technique for the throughput based on ideas from order statistics;
Section~\ref{sect:example} compares the throughput performance
of our new protocols with that of classical HARQ protocols
for the Rayleigh fading channel; 
Section~\ref{sect:conclusions} concludes the paper
and points out open questions and future work directions.

\section{System Model and Performance Metric}
\label{sect:model}

We adopt the following notation convention:
$[x]^+$ indicates $\max\{x,0\}$, 
$1_{\{x\in A\}}$ the indicator function (that equals~one whenever $x\in A$
and~zero otherwise), and
$F_X(x)=\Pr[X \leq x]$, $x\in\RR$, is the cumulative distribution function of the random variable $X$.
In the following slot, fading block and codeword length are used interchangeably.

\begin{figure}
\centering
\includegraphics[width=8cm]{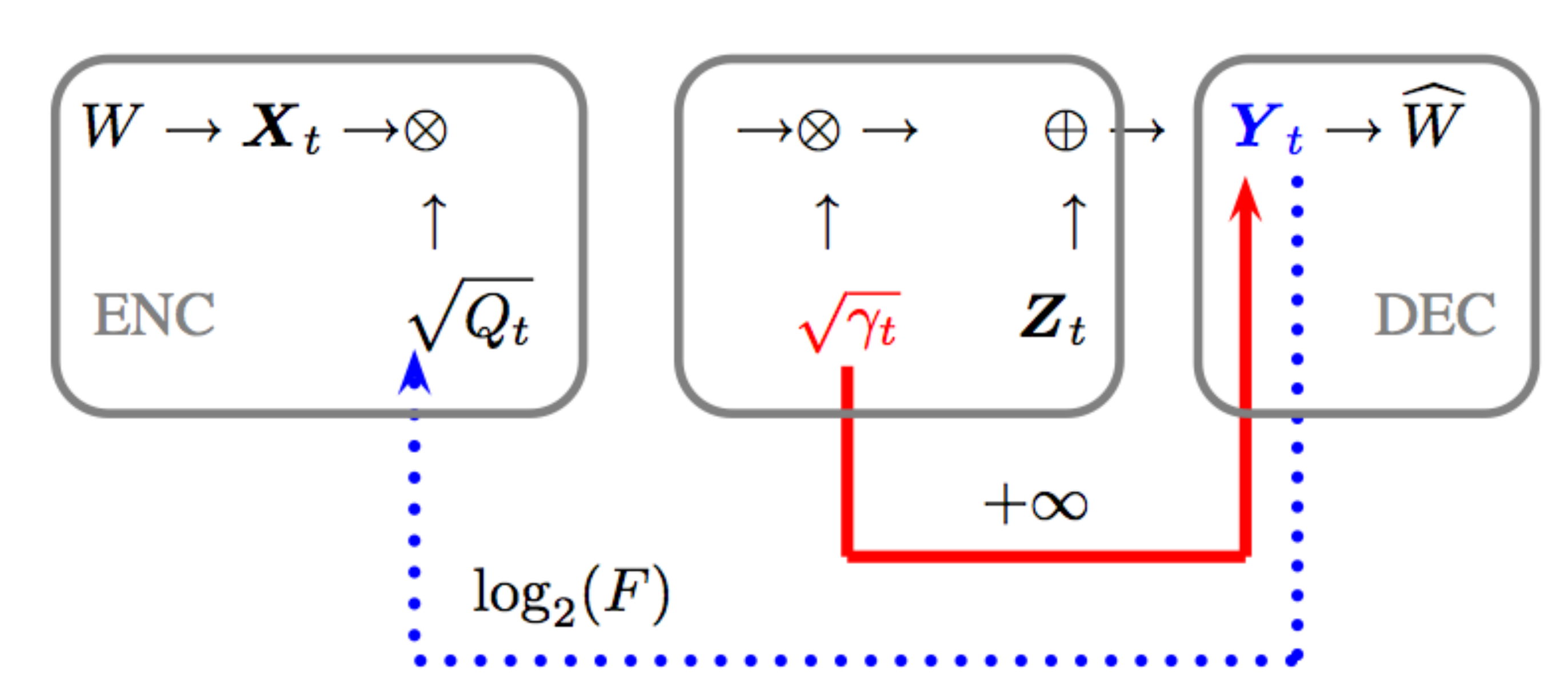}
  \caption{The block-fading Gaussian channel model with partial CSI at the encoder (dotted blue line)
  and perfect CSI at the decoder (solid red line).}
  \label{fig:GeneralModel}
\end{figure}

This work considers the single-user block-fading Gaussian channel.
The received signal vector in slot $t$, $t\in\NN$, is
\[
\Yv_t = \sqrt{\gamma_t \ Q_t} \Xv_t + \Zv_t \in\CC^L,
\]
where:
the noise $\Zv_t$ is a length-$L$ proper-complex white Gaussian random 
vector with zero mean and unit variance,
the channel fading power gain $\gamma_t$ is a scalar with $\E[\gamma_t]=1$,
the channel input signal $\Xv_t$ has Gaussian iid 
(independent and identically distributed) components 
with zero mean and unit variance, and $Q_t\geq 0$ is
the transmit power.
%
Each codeword  $\Xv_t$ spans one fading block over which the fading gain stays constant.
The slot length $L$ is such that it suffices to guarantee reliable communication if the accumulated mutual information at the receiver is above the communication rate~\cite{summa_fading_bc}. 

The fading gain $\gamma_t$ changes in an iid fashion from slot to slot.
The receiver has perfect instantaneous knowledge of $\gamma_t$ at the beginning of the slot. 
The transmitter however does not know $\gamma_t$, unless explicitly informed by the receiver.
For this reason, we assume that the transmitter cannot adjust the communication rate in each slot and thus it sends at a fixed rate.  The (partial) CSI possibly  available at the transmitter is used only for power allocation across slots.%
\footnote{We note that in practice, fading can be considered independent from slot to slot only if the
slots are separated in time by at least few channel coherence times~\cite{book:goldsmith:wireless}.
This is not a problem in multi-user systems where a user is assigned a transmission slot in every frame (and frames consist of several slots). In this case 
$\gamma_2$ indicates the fading gain of the slot where the second transmission
(first re-transmission) occurs and it needs not be the slot immediately
following the one where the first transmission occurred.
When the iid assumption does not hold, 
fading correlation across time slots can be easily incorporated in our model
by substituting products of probabilities involving different fading random
variables with the corresponding joint probabilities.}

A delay-free and error-free feedback channel with capacity $\log_2(F)$~bits per slot is available for communication of low-rate information between the receiver and the transmitter;  the receiver can feedback a retransmission request to the transmitter at the end of a slot, or quantized CSI at the beginning of a slot, or any other information representable on $\log_2(F)$~bits at any point during the slot.  We do not allow feedback bits to be accumulated over successive slots.
The case $F=1$ corresponds to absence of CSI at the transmitter,
while $F=+\infty$ corresponds to perfect CSI.

In a block-fading setting, reliable communication is possible
if the accumulated mutual information at the receiver is above the communication rate~\cite{summa_fading_bc}.
To make up for decoding errors, which occur when the channel is in deep fade, 
the transmitter can retransmit a data packet at most $M-1$~times, that is,
each data packet can be transmitted on at most $M$ channel slots.
We consider the three HARQ protocols analyzed in~\cite{caire_tuninetti:arq_it}:
\begin{itemize}
\item
{\bf ALO} (ALOha): like in slotted Aloha, the transmitter keeps sending
the same codeword and the receiver attempts decoding by using
only the most recently received codeword.
\item
{\bf RTD} (Repetition Time Diversity): the transmitter keeps sending the same
codeword and the receiver 
performs maximal ratio combining of all the received packets, thus realizing
Repetition Time Diversity.
\item
{\bf INR} (INcremental Redundancy): at each retransmission request,
the transmitter sends new redundancy
bits and the receiver optimally combines them.
\end{itemize}
The protocols work as follows.
In order to send a data packet of $b$ bits, 
the transmitter can use at most $L M$ channel uses.  For ALO and RTD, the
transmitter encodes the data packet with a Gaussian channel code of rate $b/L$
and then concatenates it with a repetition code of rate $1/M$. For INR,
the transmitter encodes the data with a Gaussian channel code of rate $b/(LM)$
and at each transmission it send a different chuck of $L$ symbols.
Following~\cite{caire_tuninetti:arq_it}, we let $R\defeq b/L$.
The throughput is defined as the {\em long-term average number of successfully decoded bits per channel use}:
\begin{equation}
\eta_{M,F}
= \lim_{T\to\infty}\frac{1}{T} 
\sum_{t=1}^{T}\,R\,\cdot\,1_{\{{\rm successful\,\,decoding\,\,in\,\,slot\,\,} t\}},
\label{eq:throughput}
\end{equation}
where the subscript $M$ indicates that at most $M$ transmissions are allowed per data packet,
and the subscript $F$ indicates that at most $F$ feedback values are allowed.
The ergodic capacity and the outage capacity are a spacial case of our framework
for $M=+\infty$ and $M=1$, respectively.
For future use, decoding fails with $m$ transmissions (and thus
a retransmission is needed) if:
\begin{subequations}
\begin{align}
      &\left\{\sum_{t=1}^{m}  \log(1+\gamma_t Q_t) <R \right\}    \ \text{for INR} \\
      &\left\{\log(1+\sum_{t=1}^{m}  \gamma_t Q_t) <R \right\}    \ \text{for RTD} \\
      &\bigcup_{t=1}^{m}\Big\{\,\log(1+\gamma_t Q_t) <R \Big\}    \ \text{for ALO}, 
\end{align}
\label{eq:pm}
\end{subequations}
since INR accumulates mutual information, RTD accumulates $\snr$, and
ALO only accounts for the most recent transmission.

In order to complete the system description, we need to specify
how the transmit power $Q_t$, $t\in\NN$, can be varied.
We assume $Q_t\in\{P_{m,f}\}$, where $P_{m,f}\geq 0$ is the power
used  at the $m$-th transmission attempt, $m\in\{1,\ldots,M\}$, 
when the feedback value is $f$, $f\in\{0,\ldots,F-1\}$.
The total transmit power must satisfy the {\em long-term average constraint~\cite{caire_taricco_biglieri:optimal_powercontrol}} defined as:
\begin{equation}
\lim_{T\to\infty}\sum_{t=1}^{T}\frac{1}{T} Q_t \leq \overline{P}, \ \text{almost surely}.
\label{eq:power}
\end{equation}
For the normalization adopted in this work, $\overline{P}$ has the meaning of average $\snr$
at the receiver. The power allocation policies are {\em causal}~\cite{negi_cioffi:dymprog_jrnl,caire_tuninetti_verdu:rate_it} in that
$Q_t$ can only depend on the (partial) knowledge of $(\gamma_1,\ldots,\gamma_t)$.

Next we show how to evaluate the throughput in~\reff{eq:throughput}
subject to the power constraint in~\reff{eq:power}.

\section{Throughput Evaluation}
\label{sect:eta eval}

In~\cite{caire_tuninetti:arq_it} we introduced a general framework
to analyze the performance of HARQ protocols based on the
{\em renewal-reward theory}~\cite{book:grimmet_strizaker:prob},
which we shall use now to evaluate the throughput.
In our system, a renewal event occurs (i.e., the system starts
anew without any memory from the past) when
the transmission of a data packet ends (either because
of successful decoding with less than $M$ transmissions
or because the maximum number of transmissions has been reached).
From~\cite{caire_tuninetti:arq_it}, the system performance
is completely characterized by the triplet
$(\Tc,\Rc,\Pc)$, where:
\begin{itemize}

\item
$\Tc\in\{1,\ldots,M\}$ in the {\em inter-renewal time}
and represents the number of slots
needed to complete the transmit of a data packet;

\item
$\Rc\in\{0,R\}$ is the {\em reward}, i.e.,
the number of bits successfully decoded per channel use
when the transmission of a data packet ends;

\item
$\Pc$ is the {\em cost}, i.e., the total transmit power
for a data packet (including all retransmissions);

\end{itemize}
Given $(\Tc,\Rc,\Pc)$, the throughput in~\reff{eq:throughput}
subject to the power constraint in~\reff{eq:power}
is given by:

\begin{theorem}[from \cite{caire_tuninetti:arq_it,elgamalcairedamen:dmtharq:it06}]
\label{thm:eta vs longtermpower}
For any $M$ and $F$
the throughput $\eta_{M,F}$ for a given power $\overline{P}$
is the solution of:
\begin{align*}
    &\eta_{M,F}^{(\star)}=
  \\&=\max \frac{\E[\Rc]}{\E[\Tc]}                 
     =\max
    \frac{R\big(1-\Pr[\Tc=M,\,{\rm failure \ to \ decode}]\big)}
         {\sum_{m=1}^{M}\Pr[\Tc\geq m]}
    \\&{\rm s.t.}\,\,\frac{\E[\Pc]}{\E[\Tc]}=
      \frac{\sum_{m=1}^{M}\E[\Pc_m|\Tc\geq m]\Pr[\Tc\geq m]}
         {\sum_{m=1}^{M}\Pr[\Tc\geq m]} \leq  \overline{P},
\end{align*}
where the maximization is over the transmit rate $R\geq 0$ and over
the power allocation
$\{\Pc_m\}_{m=1}^{M}$, where $\Pc_m\in\{P_{m,0},\ldots,P_{m,F-1}\}$
is the causal power policy for the $m$-th transmission attempt
restricted to take at most $F$ different values.
The distribution of the inter-renewal time $\Tc$ 
and the probability of failure to decode (see~\reff{eq:pm})
are function of the protocol
$\star\in\{\text{ALO, RTD, INR}\}$ used.
\end{theorem}
\begin{IEEEproof}
The proof can be found in Appendix~\ref{sec:proof thm:eta vs longtermpower}.
\end{IEEEproof}


Remarks:
\begin{enumerate}

\item
The probability of failure to decode on the last transmission, indicated as
$\Pr[\Tc=M,\,\text{failure to decode}]$, is the probability that
the data packet is lost and it is referred to as {\em outage probability}.

\item
The event $\{\Tc\leq M-1\}$ (that transmission ends before the maximum
number of transmissions has been reached) implies successful decoding.
However, a successful decoding does not necessarily imply a renewal event
(the end of the transmission of the current data packet).

\item
It is immediate to see that
the optimal power allocation meets the power constraint with equality
(otherwise, the left over power could be used on the last transmission,
which would increasing the throughput while still meeting the power constraint).

\end{enumerate}

The throughput of the different protocols for different 
values of $M$ and $F$ satisfies:
\begin{theorem}[from \cite{caire_tuninetti:arq_it}]
\label{thm:ordering of eta}
We have:
\begin{align}
\eta_{M,F}^{\rm (ALO)}\leq \eta_{M,F}^{\rm (RTD)}\leq \eta_{M,F}^{\rm (INR)}.
\label{eq:order}
\end{align}
Moreover, $\eta_{M,F}^{\rm (\star)}$ is a non-decreasing function of $M$ and of $F$,
for each protocol $\star\in\{\text{ALO, RTD, INR}\}$.
\end{theorem}
\begin{IEEEproof}
The proof of~\reff{eq:order} is as in~\cite{caire_tuninetti:arq_it} and is omitted here for sake of space. The fact that $\eta_{M,F}^{\rm (\star)}$ is a non-decreasing function of $M$ and $F$ follows by observing that if more transmissions or more accurate CSI would
hurt performance, they could just be ignored.
\end{IEEEproof}

From Theorem~\ref{thm:ordering of eta} it follows immediately that:
\begin{corollary}
For any $M$ and $F$, and for any protocol $\star\in\{\text{ALO, RTD, INR}\}$:
\[
\eta_{M=1,F}^{\rm (ALO)} \leq \eta_{M,F}^{\rm (\star)}\leq \eta_{M=\infty,F}^{\rm (INR)},
\]
where $\eta_{M=1,F}^{\rm (ALO)}$ is the {\em outage capacity} of the channel
and $\eta_{M=\infty,F}^{\rm (INR)}$ is the {\em ergodic capacity}
of the channel~\cite{caire_tuninetti:arq_it}.
\end{corollary}

In the following we first evaluate $\eta_{M=\infty,F}^{\rm (INR)}$
and $\eta_{M=1,F}^{\rm (ALO)}$ with partial CSI
and then we propose novel achievable protocols for $\eta_{M,F}^{\rm (\star)}$,
$\star\in\{\text{ALO, RTD, INR}\}$.

\section{Throughput upper bound $\eta_{M=\infty,F}^{\rm (INR)}$}
\label{sec:eta M=infty,F,(INR)}

When $M\to\infty$, the INR protocol with a time-invariant and
memoryless power allocation policy
$P_t = g(\gamma_t)$, $t\in\NN$, and with optimized rate $R$, achieves the
{\em ergodic capacity} of the channel~\cite{caire_tuninetti:arq_it} given by:
\[
\eta_{M=\infty,F}^{\rm (INR)}
= \E[\log(1+\gamma\,g(\gamma))],
\]
where the function $g(\cdot)$ can take at most $F$ different values.
In the following
we let $P_{f}\geq 0$ be the power used when the feedback
value is $f$, $f\in\{0,\ldots,F-1\}$, (i.e., we drop the index
referring to the number of repetitions, which is irrelevant here
because we considered time-invariant and
memoryless power allocation policies).
The optimal power policy $g(\cdot)$ for a finite $F$
was derived in~\cite{caireshamai:capacitypartialcsi}
and it is summarized in the following:
\begin{theorem}[from \cite{caireshamai:capacitypartialcsi}]
\label{thm:making intervals}
Let
\begin{align}
0 \leq P_0  \leq P_1 \leq  \cdots \leq P_{F-1} \leq \frac{1}{\lambda},
\label{eq: def powers for eta M=infty,F,(INR)}
\end{align}
where $P_f$ is the power used when
$\gamma\in\Rc_f^{\rm(INR)}$, with
\begin{align}
\Rc_f^{\rm(INR)}\defeq \{s_{f}  \leq \gamma  < s_{f+1}\},
\label{eq: def quantinterval for eta M=infty,F,(INR)}
\end{align}
$f\in\{0,\ldots,F-1\}$, with $s_0=0$, $s_{F}=+\infty$,   and
\begin{align}
\displaystyle
\frac{1}{s_{f}} \defeq 
\displaystyle
\frac{1}{\lambda}
\left(
\frac{\lambda(P_{f+1}-P_f)}{\eu^{\lambda(P_{f+1}-P_f)}-1}-\lambda P_f
\right),
\label{eq: def thresholds for eta M=infty,F,(INR)}
\end{align}
$f\in\{1,\ldots,F-1\}$, and where $\lambda\geq 0$ is such that 
\begin{align}
\sum_{f=0}^{F-1} P_f\,\Pr[\Rc_f^{\rm(INR)}] = \overline{P}.
\label{eq: def PC for eta M=infty,F,(INR)}
\end{align}
The ergodic capacity with partial CSI is:
\begin{align}
  &\eta_{M=\infty,F}^{\rm(INR)}=\nonumber
\\&= \max_{\{P_f\}} \sum_{f=0}^{F-1} \E\big[\log(1+\gamma P_f)| \gamma  \in \Rc_f^{\rm(INR)}\big] \Pr[\Rc_f^{\rm(INR)}],
\label{eq: def eta M=infty,F,(INR)}
\end{align}
where the maximization is subject to~\reff{eq: def powers for eta M=infty,F,(INR)}
and~\reff{eq: def PC for eta M=infty,F,(INR)}.
\end{theorem}

Remarks:
\begin{enumerate}

\item
When $F=1$ (no CSI) the transmitter can not adapt its power
across transmissions and thus sends at 
constant power $g(\gamma)=\overline{P}$.
In this case the throughput is:
\begin{align}
\eta_{M=\infty,F=1}^{\rm (INR)}
  &= \E[\log(1+\gamma\,\overline{P})].
\label{eq: def eta M=infty,F=1,(INR)}
\end{align}
Theorem~\ref{thm:making intervals} for $F=1$ gives the
result in~\reff{eq: def eta M=infty,F=1,(INR)}.

\item
With $F=\infty$ (perfect CSI) the optimal power allocation is 
water-filling~\cite{goldsmith_varaiya} given by:
\begin{align}
g(\gamma) = \left[\frac{1}{\lambda}-\frac{1}{\gamma}\right]^+,
\label{eq:def water filling}
\end{align}
and the throughput is:
\begin{align}
\eta_{M=\infty,F=\infty}^{\rm (INR)}
  &= \E\left[\left[\log\frac{\gamma}{\lambda}\right]^+\right],
\label{eq: def eta M=infty,F=infty,(INR)}
\end{align}
where the Lagrange multiplier $\lambda\geq 0$ is such that the
power constraint is met with equality.
The power policy in~\reff{eq: def thresholds for eta M=infty,F,(INR)}
for $F\to\infty$ reduces to~\reff{eq:def water filling} since 
$\frac{\lambda(P_{f+1}-P_f)}{\eu^{\lambda(P_{f+1}-P_f)}-1}\to 1$
when $P_{f+1}-P_f\to 0$ (notice that the region 
$\Rc_0^{\rm(INR)}$ always includes the interval $[0,\lambda]$
since $s_1\geq \lambda$, while the other quantization regions reduce
to a single point).

\item
The optimal quantization regions are intervals.

\item
For the purpose of numerical evaluations, it is convenient to have 
bounds on the throughput that can be fast evaluated and easily optimized. 
The throughput in Theorem~\ref{thm:making intervals} can be bounded as:
\begin{proposition}
\label{prop: bounds eta M=infty,F=infty,(INR)}
For a given set of quantization intervals 
$\Rc_f= \{s_{f}  \leq \gamma  < s_{f+1}\}$, $f\in\{0,\ldots,F-1\}$,
the ergodic capacity $\eta_{M=\infty,F}^{\rm(INR)}$
in~\reff{eq: def eta M=infty,F,(INR)} can be bounded as:
\begin{subequations}
\begin{align}
\eta_{M=\infty,F}^{\rm(INR)}
  &\!\leq\!\max_{\{s_f,P_{f}\}}\!\sum_{f=0}^{F-1} \log\big(1+\mu_f\,P_{f}\big)  \Pr[\Rc_f],
\label{eq: upper eta M=infty,F=infty,(INR)}\\
\eta_{M=\infty,F}^{\rm(INR)}
  &\!\geq\!\max_{\{s_f,P_{f}\}}\!\sum_{f=0}^{F-1} \log(1+s_{f} P_f) \Pr[\Rc_f],
\label{eq: lower eta M=infty,F=infty,(INR)}
\end{align}
\label{eq: bounds eta M=infty,F=infty,(INR)}
\end{subequations}
where $\mu_f\defeq\E[\gamma|\gamma  \in \Rc_f] \in \Rc_f$ is
the centroid of the $f$-th quantization interval, and the maximization is subject to
\begin{align}
0=s_0 \leq s_1  \leq s_2 \leq  \cdots \leq s_F= +\infty
\label{eq: how to confuse the reader with thresholds}
\end{align}
and such that the powers 
$P_{f}\geq 0$, $f\in\{0,\ldots,F-1\}$, satisfy 
$\sum_{f=0}^{F-1}P_f\Pr[\gamma  \in \Rc_f]\leq \overline{P}$.
\end{proposition}
\begin{IEEEproof}
The bounds in~\reff{eq: bounds eta M=infty,F=infty,(INR)}
follow immediately from the definition of
the quantization intervals
and from Jensen's inequality, i.e.,
\begin{align*}
  &\log(1+ \inf\{\gamma\in\Rc_f\} P_f)
\\&\leq
\E\big[\log(1+\gamma P_f)| \gamma  \in \Rc_f\big]
\\&\leq
\log(1+\E[\gamma|\gamma  \in \Rc_f] P_f),
\end{align*}
with $\inf\{\gamma\in\Rc_f\} =s_{f}$ by definition.
For both bounds in~\reff{eq: bounds eta M=infty,F=infty,(INR)}
the optimal powers are obtained by water-filling~\cite{goldsmith_varaiya}.
The optimization of the bounds in~\reff{eq: bounds eta M=infty,F=infty,(INR)}
is thus equivalent to the problem of finding the optimal quantization intervals,
which can be done efficiently by using Lloyd's algorithm~\cite{lloydalgorithm-57}.
\end{IEEEproof}

\smallskip
As the number of feedback levels $F$ increases, the quantization intervals
reduce to a single point since $\mu_f\to s_{f}$, that is,
the bounds in~\reff{eq: bounds eta M=infty,F=infty,(INR)}
converge to the water-filling ergodic capacity in~\reff{eq: def eta M=infty,F=infty,(INR)}.


\end{enumerate}

\section{Throughput Lower Bound $\eta_{M=1,F}^{\rm (ALO)}$}
\label{sec:eta M=1,F,(ALO)}

All protocols have the same throughput for $M=1$
(because retransmissions are not possible), 
which coincides with the {\em outage capacity} of the channel~\cite{caire_tuninetti:arq_it}
given by:
\[
\eta_{M=1,F}^{\rm (ALO)} = \max_{R\geq 0}
\big\{R (1-\mathsf{P}_{\rm out}(R))\big\},
\]
where $\mathsf{P}_{\rm out}(R)$ is the probability of outage given by:
\[
\mathsf{P}_{\rm out}(R) = \Pr\big[\log(1+\gamma \ g(\gamma)) < R\big],
\]
and where the non-negative function $g(\cdot)$ can take at most $F$
different values.
The optimal power allocation policy $g(\cdot)$ for a finite $F$ is:
\begin{theorem}
\label{thm:Bhashyam intervals}
Define the thresholds
\begin{align}
0\leq s_1  \leq \cdots \leq s_{F-1} \leq  s_F=s_{0} \leq s_{F+1}= +\infty,
\label{eq: def thresholds for eta M=1,F,(ALO)}
\end{align}
(notice the convention $s_F=s_{0}\in(0,\infty)$, rather than
$s_0=0$ and $s_F=+\infty$ as in~\reff{eq: how to confuse the reader with thresholds})
and the quantization regions: 
\begin{subequations}
\begin{align}
\Rc_0^{\rm(ALO)} &= \{   0  \leq \gamma  < s_1\}\cup \{ \gamma \geq  s_0\},\\
\Rc_f^{\rm(ALO)} &= \{s_{f} \leq \gamma  < s_{f+1}\}, \quad  f\in\{1,\ldots,F-1\}, 
\end{align}
\label{eq: partition for eta M=1,F,(ALO)}
\end{subequations}
Let the transmit power to be used when $\gamma\in\Rc_f^{\rm(ALO)}$ be: 
\begin{align}
P_{f}&=\frac{\eu^{R}-1}{s_f}, \ f\in\{0,\ldots,F-1\},
\label{eq: powerspowers for eta M=1,F,(ALO)}
\end{align}
The outage capacity with partial CSI is:
\begin{align}
&\eta_{M=1,F}^{\rm(ALO)}
  = \max_{\{s_f\}} 
   \Big(1-\Pr[\gamma < s_1]\Big)
\cdot\nonumber\\&\ \cdot
  \log\left(1+\frac{\overline{P}}{ 
  \sum_{f=0}^{F-1}\frac{1}{s_f}\Pr[\gamma\in\Rc_f^{\rm(ALO)}]}\right),
\label{eq: def eta M=1,F,(ALO)}
\end{align}
where the maximization is subject to the constraint in~\reff{eq: def thresholds for eta M=1,F,(ALO)}.
\end{theorem}
\begin{IEEEproof}
The proof can be found in Appendix~\ref{app:Bhashyam intervals}.

The power $P_{f}$ in~\reff{eq: powerspowers for eta M=1,F,(ALO)}
is the minimum power that guarantees no outage for all fading gains in $\Rc_f^{\rm(ALO)}$,
for $f>0$; outage can only occur when the fading gain belongs to the subset of $\Rc_0^{\rm(ALO)}$
given by $\{\gamma< s_1\}$.
\end{IEEEproof}

Remarks:
\begin{enumerate}
\item
When $F=1$ (no CSI) the transmitter can only send at constant power $g(\gamma)=\overline{P}$
and the throughput is:
\begin{align}
\eta_{M=1,F=1}^{\rm (ALO)}
  =\max_{s_1\geq 0}\log(1+\overline{P}\,s_1)\ \Pr[\gamma \geq s_1].
\label{eq: def eta M=1,F=1,(ALO)}
\end{align}
Theorem~\ref{thm:Bhashyam intervals} for $F=1$ gives the
result in~\reff{eq: def eta M=1,F=1,(ALO)}.

\item
With $F=\infty$ (perfect CSI) the optimal power allocation is
truncated channel inversion~\cite{caire_taricco_biglieri:optimal_powercontrol}:
\begin{align}
g(\gamma) = \frac{\eu^R-1}{\gamma} 1_{\{\gamma\geq s_1\}}.
\label{eq:def truncated channel inversion}
\end{align}
With~\reff{eq:def truncated channel inversion} an outage only happens 
when $\gamma< s_1$, and the throughput is
\begin{align}
\eta_{M=1,F=\infty}^{\rm (ALO)}
&= \max_{s_1\geq 0}\Pr[\gamma \geq s_1]
\log\left( 1+\frac{\overline{P}}{\E\left[\frac{1}{\gamma}\,1_{\{\gamma \geq s_1\}}\right]}\right).
\label{eq: def eta M=1,F=infinity,(ALO)}
\end{align}
When $F\to\infty$, the region
$\Rc_0^{\rm(ALO)}$ reduces to $\{0\leq \gamma  < s_1\}$ since $s_{\infty}=s_0=\infty$;
when the fading belongs to $\Rc_0^{\rm(ALO)}$ the transmit power is
$P_0=\lim_{s\to\infty}\frac{\eu^{R}-1}{s}=0$, 
thus Theorem~\ref{thm:Bhashyam intervals} reduces to truncated channel inversion.

\item
Several power allocation policies have been proposed for outage minimization
with partial CSI. However, none of the policies is optimal.
For example the solution proposed in~\cite{Bhashyam} corresponds to
the suboptimal solution $s_0=s_F=s_{F+1}=\infty$, that is,
setting the quantizations regions to be intervals.
From our result in Theorem~\ref{thm:Bhashyam intervals},
the optimal quantization regions are in general unions
of intervals.

\item
For the purpose of simplifying our numerical evaluations we propose to bound
the throughput $\eta_{M=1,F}^{\rm (ALO)}$ as:
\begin{proposition}
\label{prop: bounds eta M=1,F,(ALO)}
Let
\begin{align}
&\widehat{\eta}_{M=1,F}^{\rm (ALO)}
=\max_{0\leq s_1\leq \ldots\leq s_F\leq s_{F+1}=\infty} \Pr[\gamma \geq s_1]
\cdot\nonumber\\&\qquad \cdot
\log\left(1+\frac{\overline{P}}{\sum_{f=1}^{F}\frac{1}{s_f}\Pr[\gamma\in[s_f,s_{f+1})]}\right).
\label{eq: bound for eta M=1,F,(ALO)}
\end{align}
The throughput in~\reff{eq: def eta M=1,F,(ALO)} is bounded by
\begin{align*}
\widehat{\eta}_{M=1,F-1}^{\rm (ALO)}
\leq \eta_{M=1,F}^{\rm (ALO)}
\leq \widehat{\eta}_{M=1,F}^{\rm (ALO)}.
\end{align*}
When $F\gg1$, the optimal solution of~\reff{eq: bound for eta M=1,F,(ALO)}
tends to
\begin{align}
s_f = s_1 \xi^{f-1}, \forall f\geq 1,
\label{eq: thresholds F large for bound for eta M=1,F,(ALO)}
\end{align}
for some $\xi\geq 1$.
\end{proposition}
\begin{IEEEproof}
That $\widehat{\eta}_{M=1,F}^{\rm (ALO)}$ in~\reff{eq: bound for eta M=1,F,(ALO)}
is an upper bound for $\eta_{M=1,F}^{\rm (ALO)}$
in~\reff{eq: def eta M=1,F,(ALO)} follows by neglecting the term
$\frac{\Pr[\gamma< s_1]}{s_F}$ at the denominator in~\reff{eq: def eta M=1,F,(ALO)}.
That $\widehat{\eta}_{M=1,F-1}^{\rm (ALO)}$ in~\reff{eq: bound for eta M=1,F,(ALO)}
is an lower bound for $\eta_{M=1,F}^{\rm (ALO)}$
in~\reff{eq: def eta M=1,F,(ALO)} follows by setting $s_F=\infty$
in~\reff{eq: def eta M=1,F,(ALO)}.
It is interesting to notice that the same function $\widehat{\eta}_{M=1,F}^{\rm (ALO)}$
is a lower bound for $\eta_{M=1,F+1}^{\rm (ALO)}$ (notice the different number
of feedback values) 
and an upper bound for $\eta_{M=1,F}^{\rm (ALO)}$.

The proof of~\reff{eq: thresholds F large for bound for eta M=1,F,(ALO)}
can be found in Appendix~\ref{app:prop: bounds eta M=1,F,(ALO)}.
\end{IEEEproof}
\end{enumerate}

\section{Main result: achievable throughput for general $M$ and $F$}
\label{sect:main}
Determining $\eta_{M,F}^{(\star)}$,
$\star\in\{\text{ALO, RTD, INR}\}$,
as in Theorem~\ref{thm:eta vs longtermpower}
for general finite values of $M$ and $F$, is very complex
as it involves the solution of a dynamic program (due to the causal nature
of the power control~\cite{negi_cioffi:dymprog_jrnl,caire_tuninetti_verdu:rate_it}). 
In this section we propose novel protocols that combine repetition
and power control for a general $M$ and $F$.
The throughput of our protocols is a lower bound for the optimal
$\eta_{M,F}^{(\star)}$.

\begin{theorem}
\label{eq:our novel protocol}
For a protocol $\star\in\{\text{ALO, RTD, INR}\}$ and general finite values of  $M$ and $F$,
let $B_m\in\{0,\ldots,F-1\}$ be the feedback sent by
the receiver at the beginning of the slot corresponding to the
$m$-th transmission attempt for the current data packet, $m\in\{1,\ldots,M\}$.
Consider the following power policy: for $m\in\{1,\ldots,M\}$ 
\begin{align}
P_m =
  \frac{\eu^R-1}{\tau_{m}}\,1_{\{B_m=0\}}
+ \sum_{f=1}^{F-1}\frac{\eu^R-1}{s_{m,f}}\,1_{\{B_m=f\}},
\label{eq: def power M,F,*}
\end{align}
with
\begin{subequations}
\begin{align}
&
0 \leq \tau_m
\\&
0= s_{m,0}\leq s_{m,1} \cdots \leq s_{m,F-1}\leq s_{m,F}=+\infty.
\end{align}
\label{eq: def thresholds M,F,*}
\end{subequations}
The thresholds $\{s_{m,f}\}_{f=0}^{F}$ in~\reff{eq: def thresholds M,F,*} define
a {\em quantizer for a scaled version of the fading power gain
$\gamma_m$, where the scaling factor accounts for the information
already accumulated at the receiver in the previous $m-1$ transmissions.}
In particular, the proposed feedback policy is:
for $\star\in\{\text{ALO, RTD, INR}\}$, $m\in\{1,\ldots,M\}$ and $f\in\{0,\ldots,F-1\}$ let
\begin{subequations}
\begin{align}
&B_m = f
 \ {\rm if} \ 
   \frac{\gamma_m}{\xi_m^{(\star)}} \in[s_{m,f},  s_{m,f+1})
    \ {\rm and} \ \, \xi_m^{(\star)} > 0,
\label{eq:bm first}\\
&B_m=F-1
 \ {\rm if} \ 
\xi_m^{(\star)}\leq0,
\label{eq:bm second}
\end{align}
\label{eq:bm bm bm}
\end{subequations}
for with $\xi_1^{(\star)}=1$ and $\xi_m^{(\star)}$ for $m>1$ defined as
\begin{subequations}
\begin{align}
\xi_m^{\rm(ALO)}  &= \prod_{t=1}^{m-1}1_{\big\{1-\frac{\gamma_{t}}{\tau_{t}}>0\big\}}, 
\label{eq:xi for alo}\\
\xi_m^{\rm(RTD)}  &= 1-\sum_{t=1}^{m-1}\frac{\gamma_t}{\tau_t},
\label{eq:xi for rtd}\\
\xi_m^{\rm(INR)}  &= \left(
\frac{\eu^R}{\prod_{t=1}^{m-1}\left(1+(\eu^R-1)\frac{\gamma_t}{s_t}\right)}-1\right)\frac{1}{\eu^R-1}.
\label{eq:xi for inr}
\end{align}
\label{eq:feedback policy propsed}
\end{subequations}
The resulting throughput for protocol
$\star\in\{\text{ALO, RTD, INR}\}$,
as given in Theorem~\ref{thm:eta vs longtermpower},
is lower bounded by:
\begin{align}
&
\eta_{M,F,{\rm lb}}^{(\star)} \defeq \max_{\{\tau_m,s_{m,f}\}}
\frac{1-\mathsf{P}_{\rm out}}{1+\sum_{m=1}^{M-1}p_{m,0}}
\cdot\nonumber\\&\cdot
\log\left(1+\overline{P}\frac{1+\sum_{m=1}^{M-1}p_{m,0}}
{\sum_{m=1}^{M}\sum_{f=1}^{F-1}
\frac{p_{m,f}-p_{m,f-1}}{s_{m,f}}+\frac{p_{m,0}}{\tau_{m}}
}\right),
\label{eq: def eta M,F,* with pmf}
\end{align}
where the maximization is subject to the constraints in~\reff{eq: def thresholds M,F,*}
and where the probabilities $\{p_{m,f}\}$ are defined as:
\begin{subequations}
\begin{align}
p_{m,f} &= \widetilde{p}_{m,f}^{(\star)}       && m=1,\ldots,M, \quad f=0,\ldots,F-1, 
\label{eq:pmf  1}\\
\mathsf{P}_{\rm out} &= \widetilde{p}_{M+1,0}^{(\star)} &&(\text{by defining $s_{M+1,1}=+\infty$}),
\label{eq:pmf  3}
\end{align}
\label{eq:pmf  123}
\end{subequations}
where $\{\widetilde{p}_{m,f}^{(\star)}\}$ are defined 
in~\reff{eq:ptide alo} for ALO,
in~\reff{eq:ptide rtd} for RTD, and
in~\reff{eq:ptide inr} for INR.
\end{theorem}
The rest of the section is devoted to give a rational for the protocols in~\reff{eq:bm bm bm}-\reff{eq:feedback policy propsed}, to prove the throughput formula in~\reff{eq: def eta M,F,* with pmf} and to define the probabilities $\{\widetilde{p}_{m,f}^{(\star)}\}$ for~\reff{eq:pmf  123}.

\subsection{Protocol description and throughput evaluation}
Inspired by the power policy that minimizes the outage capacity in Theorem~\ref{thm:Bhashyam intervals},
we propose that
the transmitter uses the power policy in~\reff{eq: def power M,F,*}.
Our protocol works as follows:
\begin{itemize}
\item
The receiver feeds back $B_m=f>0$, $m\in\{1,\ldots,M\}$,
to indicate that the power
$(\eu^R-1)/s_{m,f}$ suffices to successfully decode the
current data packet when the previous $m-1$ transmissions 
are combined with the current transmission.

\item
Upon receiving $B_m=f>0$, $m\in\{1,\ldots,M\}$, the transmitter is certain that
the receiver will decode correctly with the current transmission; hence, after transmission with power
 $(\eu^R-1)/s_{m,f}$,
the transmitter prepares to sent a new data packet.

\item
The receiver feeds back $B_m=0$, $m\in\{1,\ldots,M\}$, when none of the powers 
$(\eu^R-1)/s_{m,f}$, $\forall f>0$, would guarantee successful decoding.

\item
In response to $B_m=0$, $m\in\{1,\ldots,M-1\}$, the transmitter sends with power
$(\eu^R-1)/\tau_{m}$ and prepares to retransmit the same data packet
in the next slot.

\item
In response to $B_M=0$ (for the last transmission attempt),
the transmitter sends with power
$(\eu^R-1)/\tau_{M}$ and
prepares to sent a new data packet
since no more retransmissions are permitted.
In this case an outage can occur.

\end{itemize}

In order to evaluate the throughput according to Theorem~\ref{thm:eta vs longtermpower}
we must determine the average decoded rate (reward) and
the average transmit power (cost) when the
transmission of the current data packet ends, and the average
time needed to transmit a data packet (inter-renewal time).
From the description of the protocol given above, it is clear that the
transmission of the current data packet does not end after $m$
received slots, $m\in\{1,\ldots,M-1\}$, if all the feedback values received were zero, that is, 
\begin{align*}
\Pr[\Tc\geq m] = \Pr[B_1=0,\ldots,B_{m-1}=0].
\end{align*}
When the transmission of a data packet
ends with less than $M$ transmissions, successful decoding occurs.
The transmission of the current data packet ends after
the $M$-th transmission regardless of the status of the decoder
(since no more transmission attempts are possible). 
An outage occurs if
decoding is still unsuccessful with $M$ transmissions, i.e.,
\begin{align}
\mathsf{P}_{\rm out}=\Pr[B_1=0,\ldots,B_{M}=0,\text{failure to decode}].
\label{eq: def pout M,F,*}
\end{align}
The average number of successfully decoded bits 
when the transmission of the current data packet ends,
i.e., {\em average reward}, is
\begin{align}
\E[\Rc] = R\big(1-\mathsf{P}_{\rm out}\big).
\label{eq: def ave reward M,F,*}
\end{align}
The average transmit power when transmission of the current data packet ends,
i.e., {\em average cost}, is
\begin{align}
&\E[\Pc]
=
\sum_{m=1}^{M}
\frac{\eu^R-1}{\tau_{m}}\,\Pr[B_1=0,\ldots,B_m=0]
\nonumber\\&\!\!\!\!\!\!+
\sum_{m=1}^{M}
\sum_{f=1}^{F-1}
\frac{\eu^R-1}{s_{m,f}}\,\Pr[B_1=0,\ldots,B_{m-1}=0,B_m=f].
\label{eq: def ave cost M,F,*}
\end{align}
The average time needed to transmit a data packet, i.e.,
the {\em inter-renewal time}, is
\begin{align}
\E[\Tc] 
  &= \sum_{m=1}^{M} m\ \Pr[\Tc= m]
   = \sum_{m=1}^{M}\Pr[\Tc\geq m]
\nonumber
\\&= 1 + \sum_{m=1}^{M-1}\Pr[B_1=0,\ldots,B_{m}=0].
\label{eq: def ave interenewal M,F,*}
\end{align}
%
The probabilities in~\reff{eq: def ave cost M,F,*}
and~\reff{eq: def ave interenewal M,F,*} can be easily expressed 
as a function of $\widetilde{p}_{m,f}$ defined as:
\begin{align}
\widetilde{p}_{m,f} \defeq \Pr[B_1=0,\ldots,B_{m-1}=0,B_m\leq f],
\label{eq:def widetilde pmf}
\end{align}
for  $f\in\{0,\ldots,F-1\}$
since:
\begin{align}
&\Pr[B_1=\ldots=B_{m-1}=0,B_m = f]\nonumber
\\&= \left\{\begin{array}{l l}
\widetilde{p}_{m,0} & f=0, \\
\widetilde{p}_{m,f}-\widetilde{p}_{m,f-1} & f=1,\ldots,F-2, \\
\widetilde{p}_{m-1,0} & f=F-1, \\
\end{array}\right.
\label{eq:def pmf no tilde}
\end{align}
for all $m\in\{1,\ldots,M\}$. The equality for $f=F-1$ follows since $\Pr[B_m\leq F-1]=1$ for all $m$.
As we shall proof in the next sections,
where we give the details of the protocols, 
the outage probability in~\reff{eq: def pout M,F,*}
needed for~\reff{eq: def ave reward M,F,*} ie equivalent to
$\mathsf{P}_{\rm out}=\widetilde{p}_{M+1,0}$ if we assume 
an hypothetical 1-bit of feedback at the end of the $M$-th transmission that 
indicates $B_{M+1}=1_{\{\text{failure to decode with $M$ transmissions}\}}$.
(this will correspond to a degenerate quantizer with $0=s_{M+1,0} < s_{M+1,1}=+\infty$).
This discussion justifies the definitions in~\reff{eq:pmf  123}.


Finally, with the definition in~\reff{eq:def pmf no tilde}
and by Theorem~\ref{thm:eta vs longtermpower},
the throughput for the proposed protocols is
\[
\eta= \frac{\rm eq.\reff{eq: def ave reward M,F,*}}{\rm eq.\reff{eq: def ave interenewal M,F,*}}=
{\rm eq.\reff{eq: def eta M,F,* with pmf}},
\]
where we expressed the rate $R$ in~\reff{eq: def ave reward M,F,*}
as a function of $\overline{P}$ from 
\[
\overline{P}
= \frac{\rm eq.\reff{eq: def ave cost M,F,*}}{\rm eq.\reff{eq: def ave interenewal M,F,*}}
= (\eu^R-1)
\frac{\sum_{m=1}^{M}\left(\frac{\widetilde{p}_{m,0}}{\tau_{m}}
+
\sum_{f=1}^{F-1}
\frac{\widetilde{p}_{m,f}-\widetilde{p}_{m,f-1}}{s_{m,f}}\right)}{1+\sum_{m=1}^{M}\widetilde{p}_{m,0}}.
\]

\smallskip
The probabilities $\{\widetilde{p}_{m,f}^{(\star)}\}$ in~\reff{eq:pmf  123},
as well as the feedback policy feedback policy $\{B_m\}$ in~\reff{eq:bm bm bm}
defined through the quantities $\{\xi_m^{(\star)}\}$ in~\reff{eq:feedback policy propsed},
depend on the protocol $\star\in\{\text{ALO, RTD, INR}\}$
and will be discussed next.

\subsection{First transmission (all protocols)}
In order to better understand the way the feedback value is decided,
consider the transmission on the first slot, which is the
same for all protocols.  Let $\theta\defeq\eu^R-1\geq 0$.

\begin{itemize}
\item
Feedback policy:
\begin{itemize}
\item
At the beginning of the first slot/transmission, the receiver measures $\gamma_1$ and sends
\[
B_1 = F-1  \ {\rm if} \ 
\log\left(1+\theta\frac{\gamma_1}{s_{1,F-1}}  \right)
\geq
\log(1+\theta)
\]
that is, the receiver sends back the highest possible feedback value if
the lowest possible power $\theta/s_{1,F-1}$ (from~\reff{eq: def power M,F,*}
with $m=1$ and $f=F-1$) suffices for 
successful decoding given the actual fading $\gamma_1$.
In other words, the receiver sends back
\[
B_1 = F-1 \ {\rm if} \  s_{1,F-1} \leq \gamma_1.
\]

\item
If $s_{1,F-1} > \gamma_1$,
i.e., the lowest possible power is not enough to guarantee
successful decoding, the receiver checks whether
the second lowest available power $\theta/s_{1,F-2}$
suffices for correct decoding.  The receiver sends back
\begin{align*}
B_1 = F-2
& \ {\rm if} \ 
\log\left(1+\theta\frac{\gamma_1}{s_{1,F-1}}  \right)
<
\log(1+\theta)
\\&
 \ {\rm and} \ 
\log\left(1+\theta\frac{\gamma_1}{s_{1,F-2}}  \right)
\geq
\log(1+\theta),
\end{align*}
that is,
\[
B_1 = F-2  \ {\rm if} \  s_{1,F-2} \leq \gamma_1 < s_{1,F-1}.
\]

\item
By continuing our reasoning in this manner, the receiver sends
\[
B_1 = f  \ {\rm if} \ 
\gamma_1 \in [s_{1,f}, s_{1,f+1}),
\quad f=0,\ldots,F-1,
\]
i.e., the thresholds $\{s_{1,f}\}_{f=0}^{F}$
define a quantizer for $\gamma_1$ as in~\reff{eq:bm first}
with $\xi_0=1$ (there is no scaling for the fading
value on the first transmission because there no
accumulate information at the receiver).
\end{itemize}

\item
Transmission strategy:
\begin{itemize}
\item
In response to $B_1 = f >0$ the transmitter sends with power 
$\theta/s_{1,f}$ that guarantees successful decoding
for the whole range of fading values in $[s_{1,f}, s_{1,f+1})$;
after transmission, the transmitter prepares to send a new data packet.

\item
In response to $B_1 = 0$, the transmitter sends with  
power $\theta/\tau_1$ 
\footnote{The power in this case is $\theta/\tau_1$;
notice the use of $\tau_1$ in place of $s_{1,0}$;
this is because $s_{1,0}=0$ has been already used to indicate
the left-most quantization value. The same holds for any $m\geq 1$.}
that suffices for successful
decoding only if $\gamma_1\geq \min\{\tau_1,s_{1,1}\}$;
however, the transmitter can not know whether the actual
$\gamma_1$ is above or below $\min\{\tau_1,s_{1,1}\}$,
and hence prepares to retransmit the same packet again.
\end{itemize}
\end{itemize}

From the second transmission onwards, the mode of operation
depends on the protocol used.  We will describe the
three protocols separately.

\subsection{Retransmissions for ALO}

Recall that a second transmission is triggered by $B_1=0$,
which corresponds to having sent with power $\theta/\tau_1$ on the first slot.

\begin{itemize}

\item
First retransmission:
\begin{itemize}
\item
In the ALO protocol only the most recent received slot 
is used for decoding.

Assume that in the first transmission the fading satisfied
$\gamma_1\geq \min\{\tau_1,s_{1,1}\}$.
The receiver knows that the 
transmitter will resend the same data packet in the second slot because
it received $B_1=0$ in the first slot.
The receiver can ``trick''  the transmitter into believing that it
will be able to decode in the second slot by sending $B_2=F-1$.

Clearly,
this second transmission is a waste of power, but the receiver has no other way
to inform the transmitter of its successfully decoding
owning to having already exhausted all its feedback bits at the beginning of the
current slot.
Among all possible powers that the receiver could have requested for
the second transmission, $\theta/s_{2,F-1}$  (from~\reff{eq: def power M,F,*}
with $m=2$ and $f=F-1$) is the lowest. This is captured 
in our protocol definition by the condition in~\reff{eq:bm second}.

\item
If $\gamma_1< \min\{\tau_1,s_{1,1}\}$ (which implies $\gamma_1< \tau_1$),
the receiver uses the same feedback policy it had used as on the first slot,
but with possibly different thresholds for the quantization.

Fig.~\ref{M2F2-alo} shows the feedback values for the ALO protocol with $M=2$
retransmissions and $F=2$ feedback values; the region with $B_1=B_2=0$ is divided
into two parts, the shaded region corresponds to an outage while the white
region corresponds to successful decoding.

\begin{figure}
\centering
\includegraphics[width=8cm]{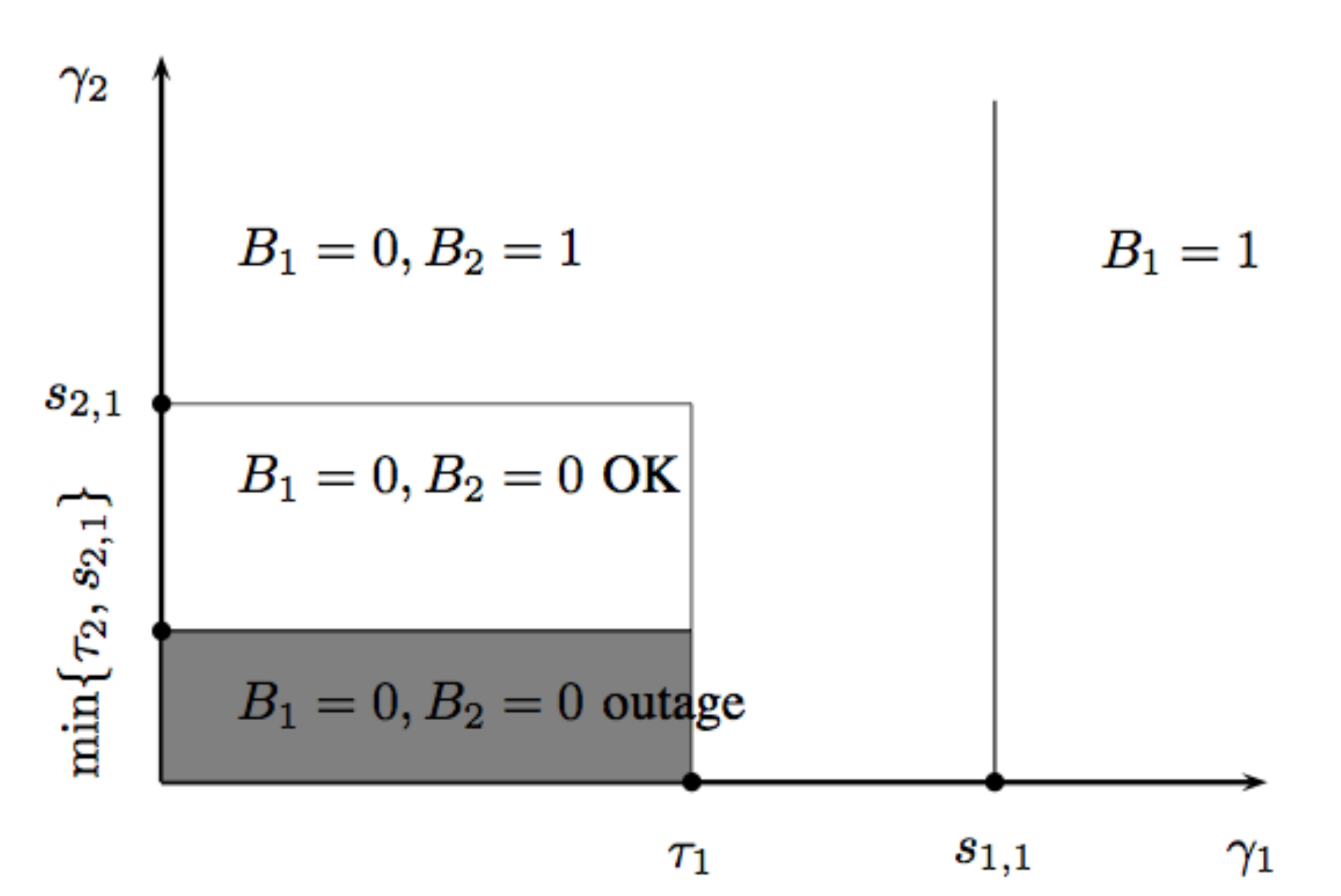}
\caption{Feedback values for the ALO protocol with $M=2$ retransmissions and $F=2$
feedback values. The shaded region corresponds to an outage.}
\label{M2F2-alo}
\end{figure}
\end{itemize}

\item
Other retransmissions:\\
By continuing our reasoning as for the first retransmission,
we arrive at the protocol definition in Proposition~\ref{eq:our novel protocol}
with the fading scaling defined in~\reff{eq:xi for alo}.

\item
Performance:\\
For ALO the probabilities $\widetilde{p}_{m,f}$, $f\in\{0,\ldots,F-2\}$, as defined in~\reff{eq:def widetilde pmf}
 are:
\begin{align}
&\widetilde{p}_{m,f}^{\rm(ALO)} \nonumber
= \Pr\left[\gamma_t < s_{t,1}, \gamma_{t-1} < \tau_{t-1} \, \forall t<m,
     \right.\nonumber\\&\left.\quad
 \gamma_{m-1} < \tau_{m-1}, \ \gamma_m < s_{m,f+1}\right]\nonumber
\\&= \Pr\left[\frac{\gamma_t}{\min\{\tau_t,s_{t,1}\}} < 1 \, \forall t<m,
  \ \gamma_m < s_{m,f+1}\right] \nonumber
\\&=\left(\prod_{t=1}^{m-1}F_\gamma(\min\{\tau_t,s_{t,1}\})\right) F_\gamma(s_{m,f+1}),
\label{eq:ptide alo}
\end{align}
with $\prod_{t=1}^{0}(\cdots) = 1$.
With the definition of $\widetilde{p}_{m,f}^{\rm(ALO)}$ in~\reff{eq:ptide alo}
we obtain the relationships in~\reff{eq:pmf  123} for the ALO protocol.

In general, the probability $\widetilde{p}_{m,f}^{\rm(ALO)}$
is available in closed form if the cumulative density function
of the fading power $F_\gamma(x)$, $x\geq 0$, is known in closed form.

\smallskip
Notice that the probability of outage (failure to decode at the last transmission)
would be equivalent to sending $B_{M+1}=0$, for this reason we have the equality
in~\reff{eq:pmf  3} by defining $s_{M+1,1}=+\infty$.
This observation holds for all protocols.
\end{itemize}

\subsection{Retransmissions for RTD}
Recall that a second transmission is triggered by $B_1=0$,
which corresponded to having sent with power $\theta/\tau_1$
on the first slot.

\begin{itemize}
\item
First retransmission:
\begin{itemize}
\item
In the RTD protocol, the receiver accumulates $\snr$.

On the second transmission, at the beginning of the slot
the receiver measures $\gamma_2$ and checks whether the
lowest available power $\theta/s_{2,F-1}$ 
suffices for decoding, and sends
\begin{align*}
  &B_2 = F-1 \ {\rm if} \
\\& 
\log\left(1+\theta\frac{\gamma_1}{\tau_1} 
              +\theta\frac{\gamma_2}{s_{2,F-1}}\right) \geq \log(1+\theta),
\end{align*}
since the resulting $\snr$  after maximal ratio combining of the two received
packets is the sum of the $\snr$  on each packet.
Hence, the feedback value at the beginning of the second slot is
\begin{align*}
B_2 = F-1  \ {\rm if} \ 
   \frac{\gamma_1}{\tau_1} 
  +\frac{\gamma_2}{s_{2,F-1}}\geq 1.
\end{align*}

\smallskip
If $\frac{\gamma_1}{\tau_1}\geq 1$, then $B_2 = F-1$.
However, if $\frac{\gamma_1}{\tau_1}\geq 1$ a retransmission 
was not necessary in the first place and
the power $\theta/s_{2,F-1}$ is wasted, as for the ALO protocol.
When ${\gamma_1}\geq {\tau_1}$ the receiver feeds back $B_2=F-1$ so that
transmission of the current data packet ends with the next slot
(with the minimum possible amount of wasted power).

\smallskip
If ${\gamma_1}\geq{\tau_1}$ and $B_2 = F-1$ then the condition
on the fading value we can rewrite as:
\begin{align*}
B_2 = F-2  \ {\rm if} \ 
   \frac{\gamma_2}{1-\frac{\gamma_1}{\tau_1}}\geq  s_{2,F-1}.
\end{align*}

\item
If instead $\frac{\gamma_1}{\tau_1} +\frac{\gamma_2}{s_{2,F-1}}< 1$
(which implies that $\frac{\gamma_1}{\tau_1}<1$)  the receiver
checks whether the second lowest available power $\theta/s_{2,F-2}$
suffices for decoding, and sends
\begin{align*}
&B_2 = F-2
\\&  \ {\rm if} \ 
  \log\left(1+\theta\frac{\gamma_1}{\tau_1} 
             +\theta\frac{\gamma_2}{s_{2,F-1}}\right) < \log(1+\theta)
\\&  \ {\rm and} \ 
\log\left(1+\theta\frac{\gamma_1}{\tau_1} 
              +\theta\frac{\gamma_2}{s_{2,F-2}}\right) \geq \log(1+\theta).
\end{align*}
Hence, the feedback value at the beginning of the second slot is
\begin{align*}
B_2 = F-2  \ {\rm if} \ 
   s_{2,F-2}\leq \frac{\gamma_2}{1-\frac{\gamma_1}{\tau_1}}<  s_{2,F-1}.
\end{align*}

\item
By proceeding with this reasoning, we see that the thresholds
$\{s_{2,f}\}_{f=0}^{F}$ define a quantizer for
$\frac{\gamma_2}{1-\frac{\gamma_1}{\tau_1}}$ when $\frac{\gamma_1}{\tau_1}< 1$.
The value $\frac{\gamma_2}{1-\frac{\gamma_1}{\tau_1}}$ to be  quantized
is larger than the actual fading $\gamma_2$ as it accounts for the $\snr$
already ``harvested'' at the receiver during the first transmission.

As an example, Fig.~\ref{M2F2-rtd} shows the feedback values for the RTD protocol with $M=2$
retransmissions and $F=2$ feedback values; the region with $B_1=B_2=0$ is divided
into two parts, the shaded region corresponds to an outage while the white
region corresponds to successful decoding.

\begin{figure}
\centering
\includegraphics[width=8cm]{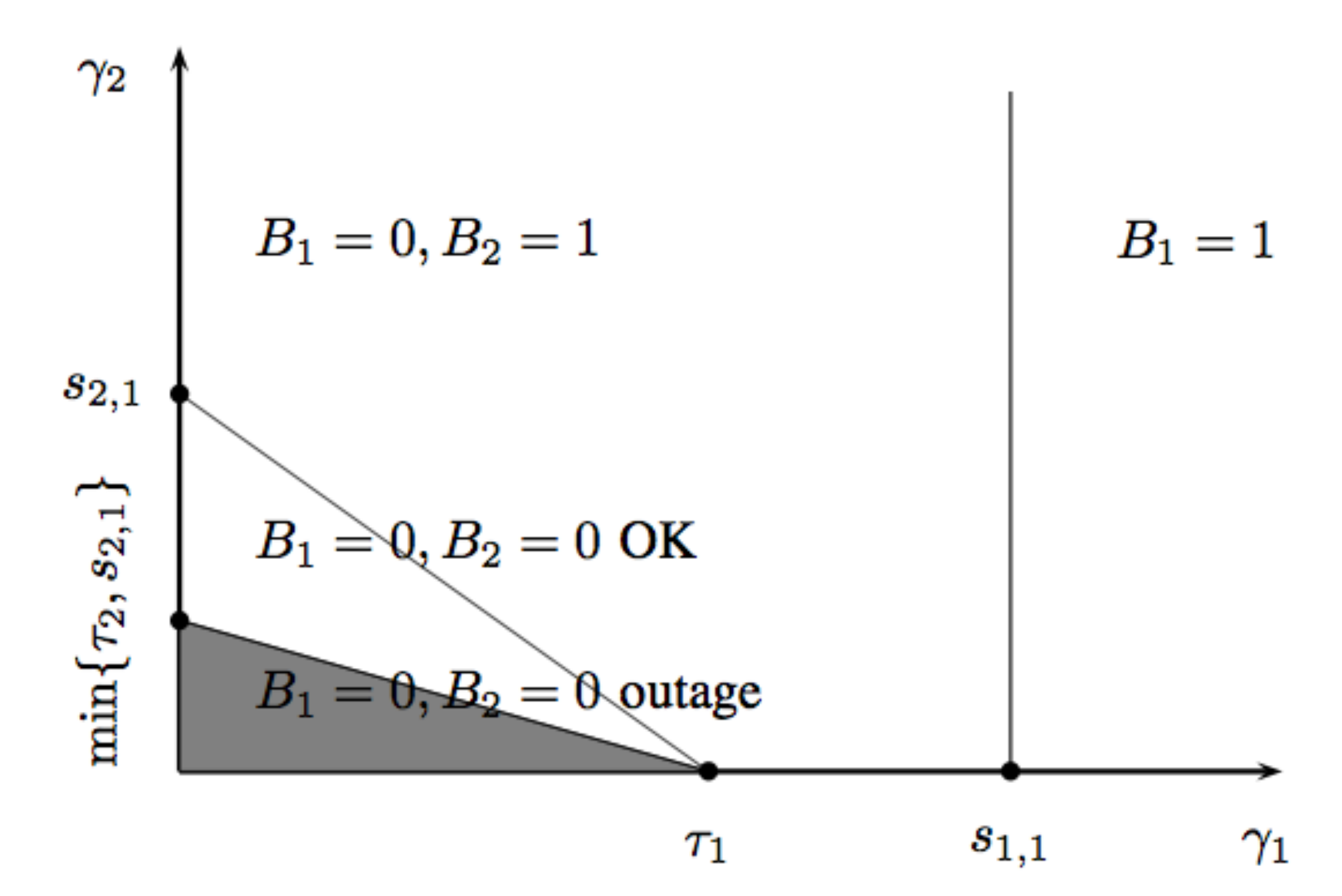}
\caption{Feedback values for the RTD protocol with $M=2$ retransmissions and $F=2$
feedback values. The shaded region corresponds to an outage.}
\label{M2F2-rtd}
\end{figure}

\end{itemize}

\item
Other retransmissions:\\
In general, if the $m$-th transmission is required, then the
receiver has already accumulated an equivalent $\snr$  of
\[
\snr'_{m-1}=\sum_{t=1}^{m-1}\frac{\gamma_t}{\tau_t},
\]
from the previous $m-1$ transmissions, all in response to a zero-value feedback value.
If $\snr'_{m-1}\geq 1$ decoding was successful, else
a retransmission is needed and the receiver uses the thresholds $\{s_{m,f}\}_{f=0}^{F}$
to define a quantizer for $\frac{\gamma_m}{1-\snr'_{m-1}}$
as proposed in Proposition~\ref{eq:our novel protocol}
(fading scaling defined in~\reff{eq:xi for rtd}).

\item
Performance:\\
For RTD the probabilities $\widetilde{p}_{m,f}$, $f\in\{0,\ldots,F-2\}$,  defined in~\reff{eq:def widetilde pmf},
are:
\begin{align}
&\widetilde{p}_{m,f}^{\rm(RTD)} \nonumber
= \Pr\left[ \frac{\gamma_t}{1-\sum_{\ell=1}^{t-1}\frac{\gamma_\ell}{\tau_\ell}}< s_{t,1},
     \right.\nonumber\\&\left.\quad
1-\sum_{\ell=1}^{t-1}\frac{\gamma_\ell}{\tau_\ell} > 0 \
 \forall t<m,
   \ \gamma_m < s_{m,f+1}\right]\nonumber
\\&= \Pr\left[\sum_{\ell=0}^{t-1}\frac{\gamma_\ell}{\tau_\ell}
     +\frac{\gamma_t}{\min\{\tau_t,s_{t,1}\}}<1 \
     \forall t<m, 
     \right.\nonumber\\&\left.\quad
     \frac{\gamma_m}{1-\sum_{\ell=0}^{m-1}\frac{\gamma_\ell}{\tau_\ell}} < s_{m,f+1}
     \right]
\label{eq:ptide rtd}
\end{align}
with $\gamma_{0}/\tau_{0}=0$.
With the definition of $\widetilde{p}_{m,f}^{\rm(INR)}$ in~\reff{eq:ptide rtd}
we obtain the relationships in~\reff{eq:pmf  123} for the RTD protocol.

In general, there is not a closed form expression available for $\widetilde{p}_{m,f}^{\rm(RTD)}$,
unless it is possible to evaluate the density of random variables of the type
$\sum_{\ell=1}^{t}\frac{\gamma_\ell}{\tau_\ell}$ in closed form.

\end{itemize}

\subsection{Retransmissions for INR}
Recall that a second transmission is triggered by $B_1=0$,
which corresponds to having sent with power $\theta/\tau_1$ on the first slot.

\begin{itemize}
\item
First retransmission:
\begin{itemize}
\item
For the INR protocol, the receiver accumulates mutual information.

At the beginning of the second slot, the receiver measures $\gamma_2$
and sends
\begin{align*}
  &B_2 = F-1
   \quad{\rm if}\quad
\\&\log\left(1+\theta\frac{\gamma_1}{\tau_1}  \right)
  +\log\left(1+\theta\frac{\gamma_2}{s_{2,F-1}}\right) \geq \log(1+\theta),
\end{align*}
since the resulting accumulated mutual information at the receiver after optimal combining
of the two transmissions is the sum of mutual information of each slot.

\smallskip
Again, if $\frac{\gamma_1}{\tau_1}\geq 1$ then $B_2=F-1$ and
a retransmission  was not necessary in the first place and
the power $\theta/s_{2,F-1}$ is wasted, as for the ALO and RTD protocols.

\smallskip
If ${\gamma_1}\geq{\tau_1}$ and $B_2 = F-1$ then the condition
on the fading value we can rewrite as:
\begin{align*}
B_2 = F-1  \ {\rm if} \ 
   \gamma_2\,\frac{1+\theta\frac{\gamma_1}{\tau_1}}{1-\frac{\gamma_1}{\tau_1}}
\geq s_{2,F-1}.
\end{align*}

\item
If $\log\left(1+\theta\frac{\gamma_1}{\tau_1}  \right)
+\log\left(1+\theta\frac{\gamma_2}{s_{2,F-1}}\right) < \log(1+\theta)$
(which implies that $\frac{\gamma_1}{\tau_1}<1$)  the receiver
checks whether the second lowest available power $\theta/s_{2,F-2}$
suffices for decoding, and sends
\begin{align*}
  &B_2 = F-2
\\&{\rm if} \
   \log\left(1+\theta\frac{\gamma_1}{\tau_1}  \right)
  +\log\left(1+\theta\frac{\gamma_2}{s_{2,F-1}}\right) < \log(1+\theta)
\\&{\rm and} \ 
   \log\left(1+\theta\frac{\gamma_1}{\tau_1}  \right)
   +\log\left(1+\theta\frac{\gamma_2}{s_{2,F-2}}\right) \geq \log(1+\theta).
\end{align*}
Hence, the feedback value at the beginning of the second slot is
\begin{align*}
&B_2 = F-2 
\ {\rm if} \ 
s_{2,F-2}\leq 
   \gamma_2\,\frac{1+\theta\frac{\gamma_1}{\tau_1}}{1-\frac{\gamma_1}{\tau_1}}
   <  s_{2,F-1}
\end{align*}

\item
By proceeding with this reasoning, we see that the thresholds
$\{s_{2,f}\}_{f=0}^{F}$ define a quantizer for
$\gamma_2\,\frac{1+\theta\frac{\gamma_1}{\tau_1}}{1-\frac{\gamma_1}{\tau_1}}$
when ${\gamma_1}< {\tau_1}$; this scaled version of $\gamma_2$ accounts for the 
mutual information already accumulated at the receiver in the first transmission.

For example, Fig.~\ref{M2F2-inr} shows the feedback values for the ALO protocol with $M=2$
retransmissions and $F=2$ feedback values; the region with $B_1=B_2=0$ is divided
into two parts, the shaded region corresponds to an outage while the white
region corresponds to successful decoding.

\begin{figure}
\centering
\includegraphics[width=8cm]{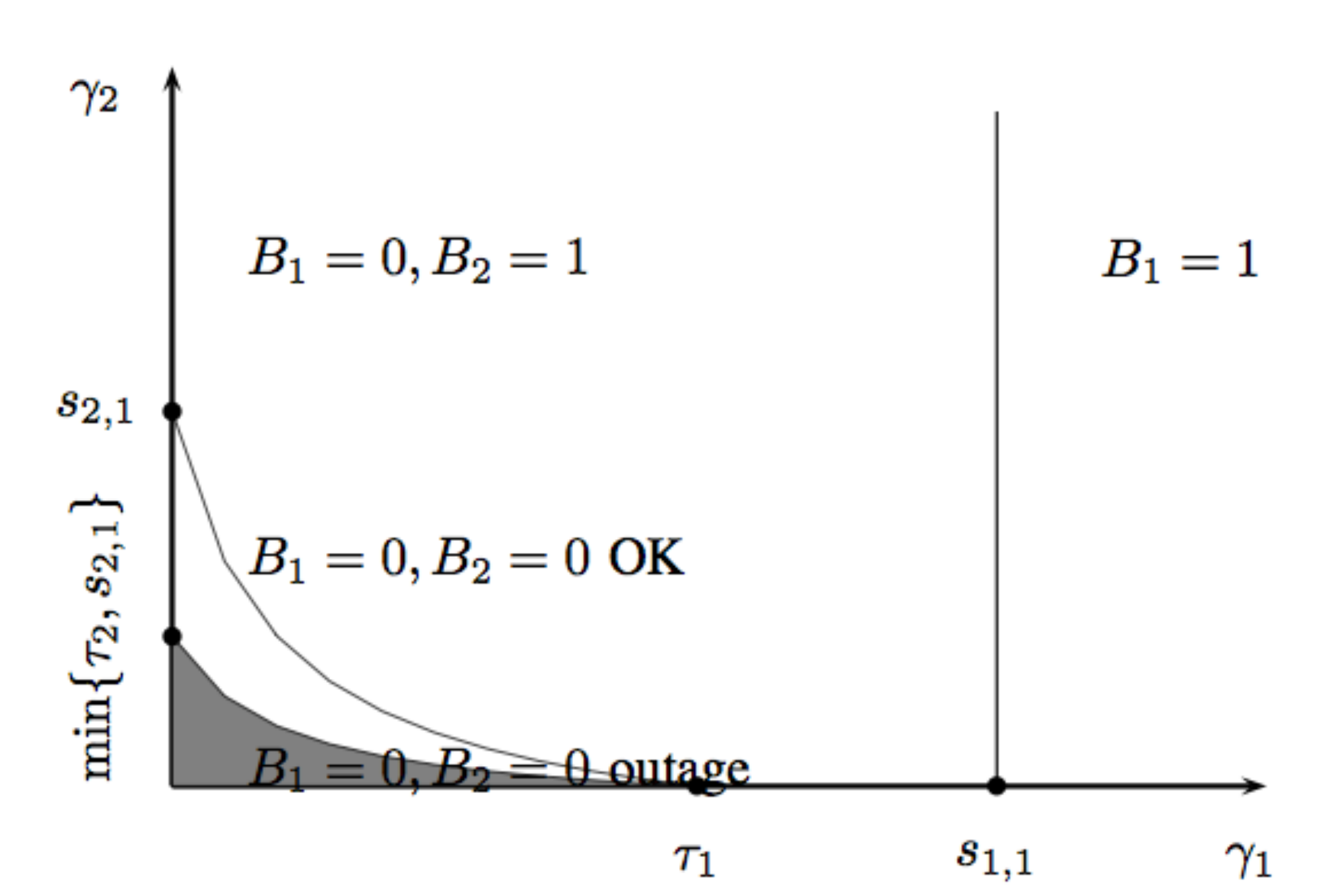}
\caption{Feedback values for the INR protocol with $M=2$ retransmissions and $F=2$
feedback values. The shaded region corresponds to an outage.}
\label{M2F2-inr}
\end{figure}

\end{itemize}

\item
Other retransmissions:\\
For a general $m$, the mutual information already accumulated from the
previous slots is
\[
\mathsf{I}_{m-1}
=\sum_{t=1}^{m-1}\log\left(1+\theta\frac{\gamma_t}{\tau_t}\right).
\]
If $\mathsf{I}_{m-1}\geq \log(1+\theta)$ then decoding was successful and the receiver sends
$B_m=F-1$ to end transmission in slot $m$.  If $\mathsf{I}_{m-1}<\log(1+\theta)$, then the receiver
looks for the smallest $f>0$ such that
\[
\mathsf{I}_{m-1} + \log\left(1+\theta\frac{\gamma_m}{s_{m,f}}\right) > \log(1+\theta).
\]
If such an $f>0$  exists, then $B_m=f$ and transmission ends with the current slot;
otherwise, $B_m=0$ and transmission continues. This procedure is equivalent 
to the protocol in Proposition~\ref{eq:our novel protocol}
with the fading scaling defined in~\reff{eq:xi for inr}.

\item
Performance:\\
For INR the probabilities $\widetilde{p}_{m,f}$, $f\in\{0,\ldots,F-2\}$, defined in~\reff{eq:def widetilde pmf},
are:
\begin{align}
&\widetilde{p}_{m,f}^{\rm(INR)} 
  = \Pr\left[
   \prod_{\ell=0}^{t-1}
   \displaystyle\frac{\left(1+\theta\frac{\gamma_\ell}{\tau_\ell}  \right)
   \left(1+\theta\frac{\gamma_t}{\min\{\tau_t,s_{t,1}\}}\right)}{1+\theta} < 1, 
   \right.\nonumber\\&\left. 
   \quad \forall t<m,\quad
   \frac{\gamma_m}{\xi_m^{\rm(INR)}} < s_{m,f+1}
    \right]
\label{eq:ptide inr}
\end{align}
with $\gamma_{0}/\tau_{0}=0$.
With this definition of $\widetilde{p}_{m,f}^{\rm(INR)}$ we obtain
the relationships in~\reff{eq:pmf  123} for the INR protocol.

In general, there is not a closed form expression available for $\widetilde{p}_{m,f}^{\rm(INR)}$,
unless it is possible to evaluate the density of random variables of the type
$\prod_{\ell=1}^{t}\left(1+\theta\frac{\gamma_\ell}{\tau_\ell}\right)$ in closed form.

\end{itemize}

\subsection{Performance with perfect CSI at the transmitter}
In order to appreciate the benefits of partial CSI at the transmitter in 
the proposed repetition protocols,
consider the performance with perfect CSI ($F=+\infty$).

Since the power control is causal, the throughput is found as a solution
of a dynamic program~\cite{book:kumar_varaiya:stochastic_systems}:
\begin{proposition}
\label{prop:our protocol f infinite}
The throughput $\eta_{M,F=\infty, {\rm lb}}^{(\star)}$ is the solution of:
\begin{subequations}
\begin{align}
\label{eq:dyprog}
&\max_{\{P_m\geq 0\}}
R\displaystyle
\frac{1-\Pr[\sum_{t=1}^{M} U_t < g(R)]}
{1+\sum_{m=1}^{M-1}\Pr[\sum_{t=1}^{m} U_t < g(R)]}
\\&
{\rm s.t.}\quad\displaystyle
\frac{\sum_{m=1}^{M}\E\left[P_m\,1_{\{\sum_{t=1}^{m-1} U_t < g(R)\}}\right]}
{1+\sum_{m=1}^{M-1}\Pr[\sum_{t=1}^{m} U_t <g(R)]}
\leq \overline{P},
\end{align}
\label{eq:dynamic prog}
\end{subequations}
where:
for ALO,
$U_m = 1_{\{\gamma_m P_m> \eu^R-1 \}}$ indicates successful 
decoding on the $m$-th transmission and $g(R)=0$;
for RTD,
$U_m = \gamma_m P_m$ represents the received $\snr$  on the $m$-th transmission and
$g(R) = \eu^R-1 \geq 0$;
%
for INR,
$U_m = \log(1+\gamma_m P_m)$
represents  the mutual information at the receiver on the $m$-th transmission and
$g(R) = R$;
the optimization is with respect to causal power policies 
$P_m=P_m(\gamma_1,\ldots,\gamma_m)$, $m=1,\ldots,M$.
\end{proposition}

%
\begin{IEEEproof}
The optimization of a causal power control system
can be cast as a {\em dynamic program over finite horizon
with complete observations}~\cite{book:kumar_varaiya:stochastic_systems,negi_cioffi:dymprog_jrnl,
caire_tuninetti_verdu:rate_it},
where the system state is $X_t = \sum_{m=0}^{t-1} U_m$,
the control is $U_t$, and the state evolves as $X_{t+1} = X_t + U_t$, with $U_0=0$,
from which~\reff{eq:dynamic prog} follows.
\end{IEEEproof}

The problem in~\reff{eq:dyprog} is similar to the outage minimization problem in~\cite{negi_cioffi:dymprog_jrnl}.
As in~\cite{negi_cioffi:dymprog_jrnl},
we can write the iterative algorithm that defines the optimal dynamic programming  solution,
however an explicit closed form solution is not available in general.  Numerical techniques, as those proposed in~\cite{negi_cioffi:dymprog_jrnl}, must be used for numerical evaluations of~\reff{eq:dynamic prog}.

\section{Novel bounding technique}
\label{sect:main bounds}
In the previous section we proposed novel protocols that
combine power control with partial CSI and retransmissions.
In all cases, a closed form expression of the throughput
requires a closed form expression for the cumulative density function of:
(a) the fading $\gamma_\ell$ for $\widetilde{p}_{m,f}^{\rm(ALO)}$
in~\reff{eq:ptide alo},
(b) random variables of the type
$S_m = \sum_{\ell=1}^{m}\frac{\gamma_\ell}{\tau_\ell}$,
for $\widetilde{p}_{m,f}^{\rm(RDT)}$ in~\reff{eq:ptide rtd}, and
(c) random variables of the type
$I_m = \prod_{\ell=1}^{m}\left(1+\theta\frac{\gamma_\ell}{\tau_\ell}\right)$
for some $\theta\geq 0$
for $\widetilde{p}_{m,f}^{\rm(INR)}$ in~\reff{eq:ptide inr}.
Since the distribution of $I_m$ is rarely known in closed form,
in the following we propose a novel bounding technique for the
cumulative density function of $I_m$
in terms of the the cumulative density function of $S_m$.
For the case of iid Rayleigh fading, the density of $S_m$
is known in closed form; hence, our technique allows to
determine closed-form upper and lower bounds
for probabilities involving $I_m$.

Consider generic non-negative constants $\{\tau_m\}$, $m\in\NN$,
a constant $\theta\geq 0$, and define
\begin{subequations}
\begin{align}
p_m^{\rm (ALO)}
  &= \Pr\left[\frac{\gamma_s}{\tau_s} < 1 , \, s=1,\ldots,m\right],
\label{eq:pmalo}
\\
p_m^{\rm (RTD)}
  &= \Pr\left[\sum_{s=1}^{m}\frac{\gamma_s}{\tau_s} < 1 \right],
\label{eq:pmrtd}
\\
p_m^{\rm (INR)}
   &= \Pr\left[\frac{1}{\log(1+\theta)}\sum_{s=1}^{m}\log\left(1+\theta\frac{\gamma_s}{\tau_s}\right) < 1 \right], 
\label{eq:pminr}
\end{align}
\label{eq:pm all}
\end{subequations}
for some sequence $\{\gamma_\ell\}$ of iid random variables.

\begin{figure}
\centerline{\includegraphics[width=3.5in]{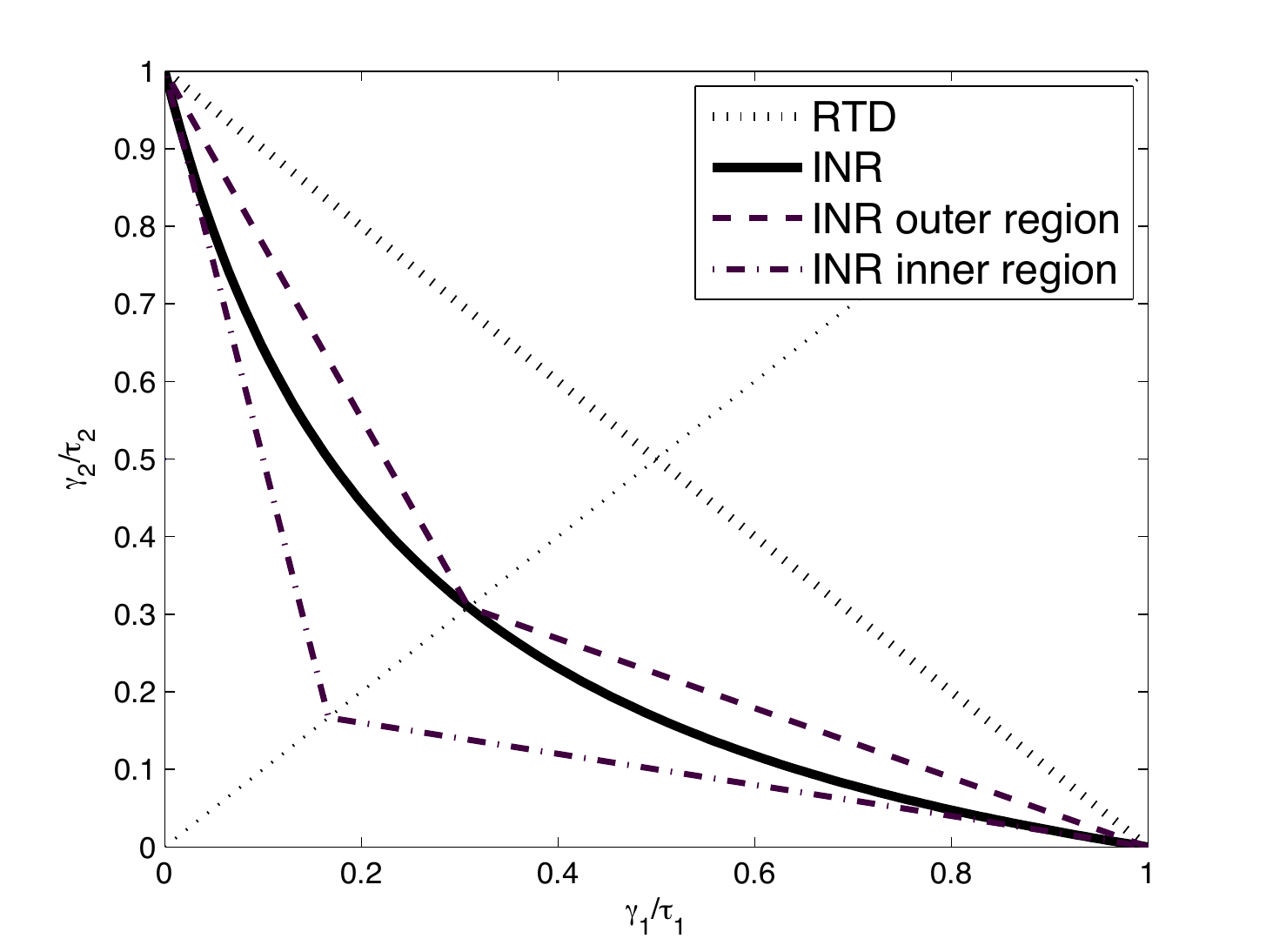}}
\caption{The region that defines the probability of outage for INR
with $M=2$ and its approximations.}
\label{INRregion}
\end{figure}

{\bf Example:}
As an example, consider Fig.~\ref{INRregion}, which shows in the plane
$x_1\defeq \frac{\gamma_1}{\tau_1},x_2\defeq \frac{\gamma_2}{\tau_2}$
the regions that defines $p_2^{(\star)}$, $\star\in\{\text{ALO, RTD, INR}\}$.
The probability $p_2^{\rm (ALO)}$
is the integral of the joint density of $(x_1,x_2)$ over the square 
$(x_1,x_2)\in[0,1]^2$.   The probability $p_2^{\rm (RTD)}$
is the integral over the triangle $x_1+x_2\leq 1$,
$x_1\geq 0$ and $x_2\geq 0$, that is, over the region in the
positive quadrant below the dotted-line curve labeled ``RTD'' in Fig.~\ref{INRregion}.
And finally, the probability $p_2^{\rm (INR)}$ is the integral over
the region in the positive quadrant below the solid-line curve labeled ``INR''
in Fig.~\ref{INRregion}.
The curve labeled ``INR'' in Fig.~\ref{INRregion}
is a convex function that can be bounded from
above and from below by piece-wise linear functions.
We chose piece-wise linear
functions because the region they define is the
union of triangular regions.
%
In particular, for the inner bound, 
we take the union of the two regions below the tangent lines at 
$(x_1,x_2)=(0,1)$ and at $(x_1,x_2)=(1,0)$ (the region in the
positive quadrant below the dash-dotted-line
curve labeled ``INR inner region''
in Fig.~\ref{INRregion}), while for the outer bound, 
we take the union of the two regions below the lines passing through
$(x_1,x_2)=(0,1)$ and $(x_1,x_2)=(1/2,1/2)$, and through
$(x_1,x_2)=(1,0)$ and $(x_1,x_2)=(1/2,1/2)$ (the region in the
positive quadrant below the dashed-line
curve labeled ``INR outer region''
in Fig.~\ref{INRregion}).

By extending the idea presented in the above example
to the case of a general $m\in\NN$ we can show:
\begin{proposition}
\label{prop:p  INR m bounds}
The probability in~\reff{eq:pminr} can be bounded as:
\begin{subequations}
\begin{align}
p_m^{\rm (INR)}
  &\geq
\Pr\left[ 
 (1+\theta) \sum_{s=1}^{m}\frac{\gamma_s}{\tau_s}
 -\theta \, \max_{t=1,\ldots,m} \frac{\gamma_t}{\tau_t}
 < 1
  \right]
\label{eq:p  INR m geq}
%
\\
p_m^{\rm (INR)}
  &\leq
\Pr\left[ 
 \sum_{s=1}^{m}\frac{\gamma_s}{\tau_s}+
 \right.\nonumber\\&\left.\quad
 + \left(\frac{\theta}{(1+\theta)^{1/m}-1}-m\right) \max_{t=1,\ldots,m} \frac{\gamma_t}{\tau_t}
 < 1
  \right]
\label{eq:p  INR m leq}
\end{align}
\label{eq:p  INR m bounds}
\end{subequations}
\end{proposition}
\begin{IEEEproof}
The proof can be found in Appendix~\ref{app:p  INR m bounds}.
\end{IEEEproof}

The interesting fact about the two bounds in~\reff{eq:p  INR m bounds}
is that they are computable from the knowledge of the density of the random variable
\begin{align}
\label{eq:xab}
X_{a,b} = 
a \,\max_{t=1,\ldots,m} \frac{\gamma_t}{\tau_t} +
 b\,\sum_{s=1}^{m}\frac{\gamma_s}{\tau_s} 
\end{align}
for some fixed $(a,b)\in\RR^2$.


\begin{proposition}
\label{prop:density of sumt gammat/taut}
For the case of iid negative exponential
random variables $\{\gamma_t\}$ (i.e., iid Rayleigh fading),
the density of $X_{a,b}$ in~\reff{eq:xab},
for any $(a,b)\in\RR^2$, is given in~\reff{eq:finally out pfd!}
in Appendix~\ref{app:orderstst}.
\end{proposition}
\begin{IEEEproof}
The proof can be found in Appendix~\ref{app:orderstst}.
\end{IEEEproof}

\section{The iid Rayleigh fading channel}
\label{sect:example}

\begin{figure}
\centerline{\includegraphics[width=3.5in]{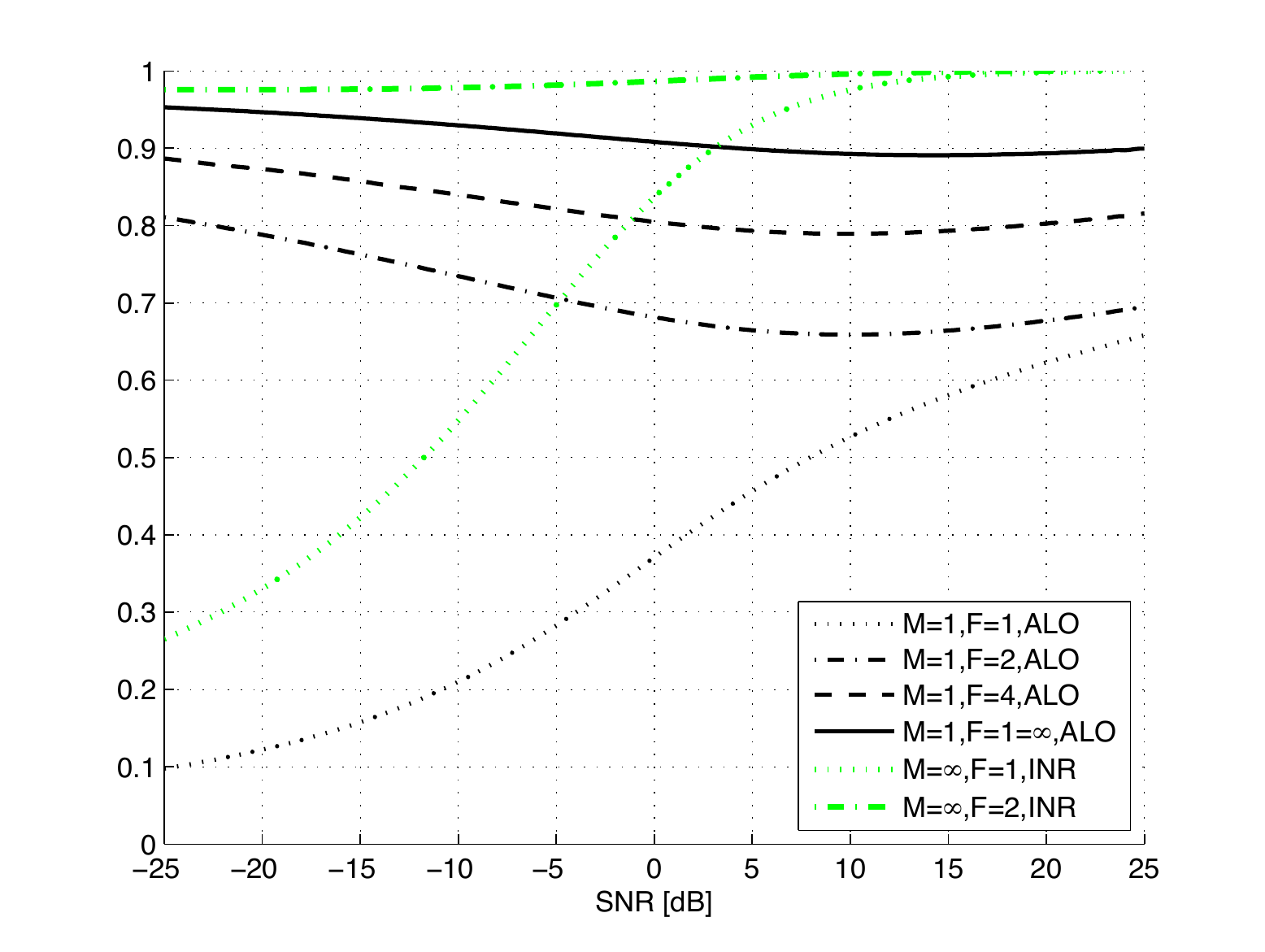}}
\caption{Ratio between the throughput
and  $\eta_{M=\infty,F=\infty}^{\rm(INR)}$ (the ergodic capacity with full CSI),
for the Rayleigh fading channel.}
\label{fig:thrratio}
\end{figure}

\begin{figure}
\centerline{\includegraphics[width=3.5in]{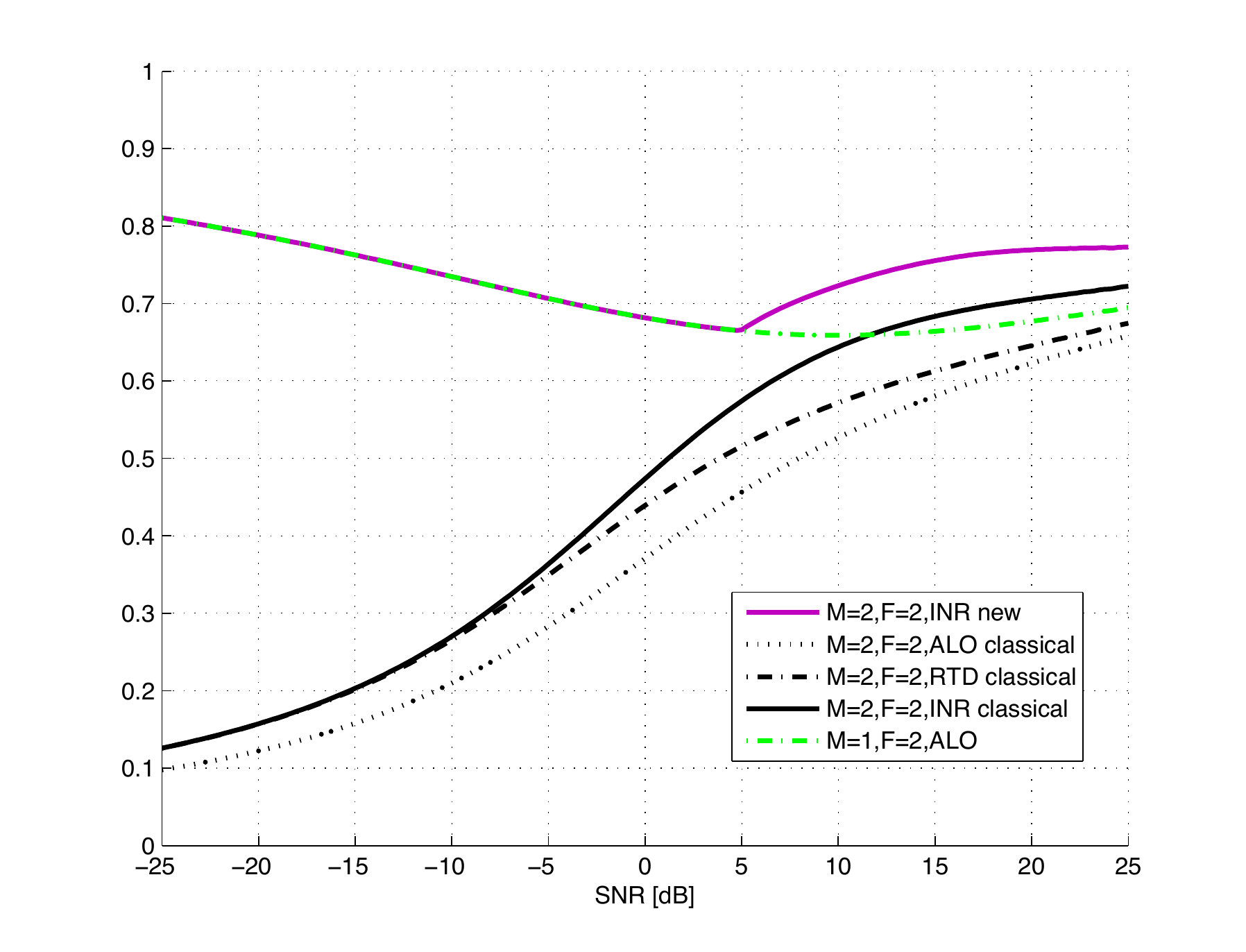}}
\caption{Ratio between the throughput
and $\eta_{M=\infty,F=\infty}^{\rm(INR)}$ (the ergodic capacity with full CSI),
for the Rayleigh fading channel.}
\label{fig:thrratioszoom}
\end{figure}

To illustrate the gain achievable with the 
protocols  proposed in Section~\ref{sect:main},
we evaluate the performance of the different protocols
for the Gaussian iid Rayleigh fading channel,
for which the fading cumulative distribution function is
$F_\gamma(x) = 1-\eu^{-x}$ for $x\geq 0$.
We define the exponential integral function as:
\[
\E\left[\frac{1}{\gamma}\, 1_{\{\gamma\geq x\}}\right]
\stackrel{\rm RF}{=}\int_{x}^{\infty}\eu^{-t}/t\,\diff{t}
\defeq {\rm Ei}(x)
\] for $x\geq 0$.
We use the symbol ``$\stackrel{\rm RF}{=}$'' to indicate
that the equality holds for Rayleigh fading channels.

For the plots, communication rates are measured in bits/sec/Hz
and the figures show the relative throughput performance with
respect to the ergodic water-filling capacity (i.e., $M=F=\infty$ and INR),
which is the ultimate performance limit for a fading channels with full CSI.
Table~\ref{tab:summary} summarizes the cases considered in the following.

\begin{center}
\begin{table}[!h]
\caption{}
\label{tab:summary}
\begin{tabular}{|c|l|l|l|}
\hline
Amount  & $M=1$                      &  $M=2$                & $M=\infty$ \\
of CSI  & outage cap.                &  HARQ protocols       & ergodic cap. \\
\hline
Absent  & in~\ref{ray M=1 F=1 ALO}   & (impossible, need at & in~\ref{ray M=inf F=1 INR} \\
($F=1$) &                            &  least 1-bit for ack/nack) &                       \\
\hline
1 bit   & in~\ref{ray M=1 F=2 ALO}   & in~\ref{ray M=2 F=2 old} 
                                               (classical HARQ)& in~\ref{ray M=inf F=2 INR} \\
($F=2$) &                            & in~\ref{ray M=2 F=2 new} 
                                                (proposed HARQ)&                       \\
\hline
Full    &  in~\ref{ray M=1 F=inf ALO}& (not evaluated)       & in~\ref{ray M=inf F=inf INR} \\
($F=\infty$) &                       &                       &                       \\
\hline 
\end{tabular}
\end{table}
\end{center}

\subsection{Throughput upper bound (ergodic capacity)}

\subsubsection{$M=\infty$, $F=1$, INR}
\label{ray M=inf F=1 INR}
With constant power allocation the ergodic capacity is:
\begin{align*}
\eta_{M=\infty,F=1}^{\rm (INR)}
\stackrel{\rm RF}{=} \eu^{\frac{1}{\overline{P}}}{\rm Ei}\left(\frac{1}{\overline{P}}\right).
\end{align*}

\subsubsection{$M=\infty$, finite $F$, INR}
\label{ray M=inf F=2 INR}
The ergodic capacity with partial CSI is given in Theorem~\ref{thm:making intervals}:
\begin{align*}
&\eta_{M=\infty,F}^{\rm(INR)}
\\&\stackrel{\rm RF}{=}  \max 
\Big\{\sum_{f=0}^{F-1}
-\log(1+P_f s_{f}) \eu^{-s_{f}} -\eu^{1/P_f} {\rm Ei}(1/P_f +s_{f})
 \\&
+\log(1+P_f s_{f-1})     \eu^{-s_{f-1}}     +\eu^{1/P_f} {\rm Ei}(1/P_f+ s_{f-1})
\Big\}
\end{align*}
where the thresholds $\{s_f\}$ are defined in~\reff{eq: def thresholds for eta M=infty,F,(INR)},
the maximization over $\{P_f\}$
is subject to the constraint in~\reff{eq: def powers for eta M=infty,F,(INR)} and 
\begin{align*}
\overline{P}
\stackrel{\rm RF}{=} \sum_{f=0}^{F-1} P_f [\eu^{-s_{f-1}}-\eu^{-s_{f}}]
\end{align*}

The ergodic capacity with partial CSI can be bounded by using
Proposition~\ref{prop: bounds eta M=infty,F=infty,(INR)};
in particular, $\mu_f$ in~\reff{eq: upper eta M=infty,F=infty,(INR)}
is given by
\begin{align*}
\mu_f 
\stackrel{\rm RF}{=}
\frac{\eu^{-s_{f-1}}(1+s_{f-1})-\eu^{-s_{f}}(1+s_{f})}{\eu^{-s_{f-1}}-\eu^{-s_{f}}}.
\end{align*}

\subsubsection{$M=\infty$, $F=\infty$, INR}
\label{ray M=inf F=inf INR}
With water-filling power allocation the ergodic capacity is:
\begin{align*}
\eta_{M=\infty,F=\infty}^{\rm (INR)}
\stackrel{\rm RF}{=} {\rm Ei}(\lambda),
\\
\overline{P}
\stackrel{\rm RF}{=} \frac{\eu^{-\lambda}}{\lambda}-{\rm Ei}(\lambda).
\end{align*}

\subsubsection{Discussion}

From Fig.~\ref{fig:thrratio}
we see that $\eta_{M=\infty,F=2}^{\rm (INR)}$ 
(i.e., 1 bit of feedback) is 
already at 97\% of the water-filling capacity
$\eta_{M=\infty,F=\infty}^{\rm (INR)}$
at an $\snr$  as low as -25dB.
At high $\snr$,
$\eta_{M=\infty,F=1}^{\rm (INR)}$ (no CSI)
behaves like $\eta_{M=\infty,F=\infty}^{\rm (INR)}$ (full CSI)
but at low $\snr$  their ratio tends to zero. 
This restates a well know fact that 
power allocation in single user channels offers benefits at low $\snr$  only
and that a single bit of feedback (i.e., $F=2$)
gives almost all the gain achievable by full CSI  (i.e., $F=\infty$).

\subsection{Throughput lower bound (outage capacity)}

\subsubsection{$M=1$, $F=1$, ALO}
\label{ray M=1 F=1 ALO}

With constant power allocation the outage capacity is:
\begin{align*}
\eta_{M=1,F=1}^{\rm (ALO)}
\stackrel{\rm RF}{=} \max_{s_1\geq 0} \log(1+\overline{P}s_1)\eu^{-s_1}.
\end{align*}

\subsubsection{$M=1$, $F=2$, ALO}
\label{ray M=1 F=2 ALO}
With 1-bit of feedback, the throughput is the solution of:
\begin{align*}
&\eta_{M=1,F=2}^{\rm (ALO)}
  = \max_{0\leq s_1\leq s_2\leq \infty}
\eu^{-s_1}
\cdot\\&\qquad \cdot
\log\left(1+\frac{\overline{P}}{
 \frac{1-\eu^{-s_1}}{s_2}
+\frac{\eu^{-s_1}-\eu^{-s_2}}{s_1}
+\frac{\eu^{-s_2}}{s_2}
}  \right)
\\&= \max_{0\leq s_1\leq \infty}\eu^{-s_1}\log(1+\overline{P}s_1\eu^{s_1}).
\end{align*}

\subsubsection{$M=1$, $F=\infty$, ALO}
\label{ray M=1 F=inf ALO}
With truncated channel inversion power allocation the outage capacity is:
\begin{align*}
\eta_{M=1,F=\infty}^{\rm (ALO)}
\stackrel{\rm RF}{=} \max_{\tau\geq 0} \log\left(1+\frac{\overline{P}}{{\rm Ei}(\tau)}\right)\eu^{-\tau}.
\end{align*}

\subsubsection{Discussion}
When $M=1$, there is no need to send a ACK/NACK
because the transmitter cannot retransmit.
In this case, the one bit of feedback should
indeed be used at the beginning of the slot to inform the transmitter
about the state of the channel. From Fig.~\ref{fig:thrratio}
we observe that 1-bit of feedback at -25dB results in
$\eta_{M=1,F=2}^{\rm (ALO)}/\eta_{M=\infty,F=\infty}^{\rm (INR)}= 81\%$
while (with constant power) $\eta_{M=1,F=1}^{\rm (ALO)}/\eta_{M=\infty,F=\infty}^{\rm (INR)}= 10\%$ only.
At +5dB, $\eta_{M=1,F=2}^{\rm (ALO)}/\eta_{M=\infty,F=\infty}^{\rm (INR)}= 73\%$
while (with constant power) $\eta_{M=1,F=1}^{\rm (ALO)}/\eta_{M=\infty,F=\infty}^{\rm (INR)}= 45\%$.
In fact, at high $\snr$, power allocation is less critical and
the gain due to CSI vs. no CSI diminishes.
%
We also reported for comparison the achievable throughput 
for ALO with $F=4$ (2 bits of feedback)
by using the approximation  $s_\ell = s_1\,\xi^{\ell-1}$
as in Proposition~\ref{prop: bounds eta M=1,F,(ALO)},
and for ALO with $F=\infty$ (full CSI).
We see that the gains attainable at low $\snr$  due to only a few bits of feedback are dramatic
and that 2 bits of feedback attain a throughput remarkably close to the case with full CSI.

\subsection{Classical HARQ protocols with $F=2$ and $M=2$}
\label{ray M=2 F=2 old}

Classical HARQ protocols use the 1 bit of feedback (i.e., $F=2$) to signal ACK/NACK.
Classical HARQ protocols are a special case of the protocols proposed
in Section~\ref{sect:main} obtained by setting $s_{m,f}=\infty$ for all $m>0$ and $f>0$ in~\reff{eq: def thresholds M,F,*}.
In classical HARQ protocols the power can vary across repetitions (i.e., different values
of $\tau_m$, $m=1,\ldots,M$) but it cannot depend on the CSI~\cite{elgamalcairedamen:dmtharq:it06}.
In this case the throughput in~\reff{eq: def eta M,F,* with pmf}
is a function only of $\{\widetilde{p}_{m,0}\}$, $m\in\{1,\ldots,M+1\}$ given by:
%
%
\begin{align*}
\widetilde{p}_{m,0}^{\rm (ALO)}
  &\stackrel{\rm RF}{=} \prod_{s=1}^{m}(1-\eu^{-\tau_s})
\\
\widetilde{p}_{m,0}^{\rm (RTD)}
  &\stackrel{\rm RF}{=}
\sum_{s=1}^{m}\frac{1-\eu^{-\tau_s}}{\prod_{s\not= j}(1-\tau_s/\tau_j)}
\\
\widetilde{p}_{m,0}^{\rm (INR)}
  &= \Pr\left[\sum_{s=1}^{m}\log\left(1+\theta\frac{\gamma_s}{\tau_s}\right) < \log(1+\theta) \right],
\end{align*}
for $\theta\defeq \eu^R-1 \geq 0$ and $m\geq 1$.
%
Consider the case $M=2$: the only probability not known in closed form is
$\widetilde{p}_{m,0}^{\rm (INR)}$, which we bound
by using the technique developed
in Proposition~\ref{prop:density of sumt gammat/taut} in Section~\ref{sect:main bounds} as follows:
\begin{align*}
     \widetilde{p}_{2,0}^{\rm (in)}  &= 1- q\left(\frac{1}{2+\theta}\right)
\\
\leq \widetilde{p}_{2,0}^{\rm (INR)} &=\E\left[F_\gamma\left(\tau_2\frac{1-\gamma_1/\tau_1}{1+\gamma_1/\tau_1\,\theta}\right)\right]
\\
\leq \widetilde{p}_{2,0}^{\rm (out)} &= 1- q\left(\frac{\sqrt{1+\theta}-1}{\theta}\right)
\\
\leq \widetilde{p}_{2,0}^{\rm (RTD)} &= 1 - q(1/2)
= 1-\frac{\eu^{-\tau_1}}{1-\tau_1/\tau_2}-\frac{\eu^{-\tau_2}}{1-\tau_2/\tau_1}
\\
\leq \widetilde{p}_{2,0}^{\rm (ALO)} &=(1-\eu^{-\tau_1})(1-\eu^{-\tau_2})
\end{align*}
where the function $q(x)$ is given by~\reff{eq:finally out pfd!} in Appendix~\ref{app:orderstst}.

\subsubsection{Discussion}
In Fig.~\ref{fig:thrratioszoom}, classical HARQ protocols are labeled as ``classical''
in order to distinguish them from the novel protocols proposed in this work, 
which are labeled as ``new''.  The case of $M=2$ transmission is considered
(which implies that the throughput needs to be optimized with respect to the
two parameters $(\tau_2,\tau_1)$). From the numerical results for
$\snr=\overline{P}\in[-25,+25]$dB, we saw that equal power allocation 
across transmissions (i.e., $\tau_2=\tau_1$) is optimal at all $\snr$'s for ALO; 
for RTD and INR, the use of different power (i.e., $\tau_2\not=\tau_1$)
offers benefits; however the improvement is negligible
(for example less than 0.3\% across the entire range of simulated powers for RTD)
for this reason in Fig.~\ref{fig:thrratioszoom} we only show the throughput
with the optimized $(\tau_2,\tau_1)$ only for INR.
We see that the throughput trend for the classical HARQ protocols
with $F=2$ and $M=2$ is the same as the one for $F=1$ and $M=1$ 
reported in Fig.~\ref{fig:thrratio}. This shows that the 1 bit of feedback 
used for ACK/NACK only does not offer substantial throughput improvement
compared to the outage capacity with constant power allocation ($F=1$ and $M=1$).
Only for $\snr$'s larger than 10dB, using classical INR with $F=2$ and $M=2$ 
gives a larger throughput that ALO with $F=1$ and $M=1$ with power control.
From these observations we conclude that classical HARQ make
an inefficient use of the feedback resources.

\subsection{New protocols for $M=2$ and $F=2$}
\label{ray M=2 F=2 new}
Since we consider the case $F=2$, we only need to characterize
$\widetilde{p}_{m,f}$, for $m\in\{1,2,3\}$ and $f=0$,
as per Proposition~\ref{eq:our novel protocol}.

The probability of requesting a retransmission is:
\begin{align*}
\widetilde{p}_{1,0}
&=\Pr[B_1=0]
=\Pr\left[\frac{\gamma_1}{s_{1,1}}< 1 \right]
 =\Pr[\Tc=2]
\\&=\widetilde{p}_{2,1};
\end{align*}
the probability of decoding failure after the first transmission is:
\begin{align*}
&\widetilde{p}_{2,0}=\Pr[B_1=0,B_2=0]=\\
&\Pr\left[\frac{\gamma_1}{\min\{\tau_1,s_{1,1}\}}< 1,
   \frac{\gamma_2}{s_{2,1}}< 1\right]
   & {\rm ALO} \\  
&\Pr\left[\frac{\gamma_1}{\min\{\tau_1,\tau_{1,1}\}}< 1,
   \frac{\gamma_1}{\tau_{1}}+
   \frac{\gamma_2}{s_{2,1}}< 1\right]
   & {\rm RTD} \\  
&\Pr\left[\frac{\gamma_1}{\min\{\tau_1,\tau_{1,1}\}}< 1,
    \frac{\gamma_1}{\tau_{1}}+
    \frac{\gamma_2}{s_{2,1}}\Big(1+
    \theta  \frac{\gamma_1}{\tau_{1}}\Big)< 1\right] 
   & {\rm INR},
\end{align*}
and the probability of decoding failure after the second transmission,
which coincides with the probability of outage, is:
\begin{align*}
&\widetilde{p}_{3,0}=\Pr[B_1=0,B_2=0,B_3=0]=\\
&\Pr\left[\frac{\gamma_1}{\min\{\tau_1,s_{1,1}\}}\!<\!1,
   \frac{\gamma_2}{\min\{\tau_{2},s_{2,1}\}}\!<\!1\right]
   \quad {\rm ALO} \\  
&\Pr\left[\frac{\gamma_1}{\min\{\tau_1,s_{1,1}\}}\!<\!1,
   \frac{\gamma_1}{\tau_{1}}+\frac{\gamma_2}{\min\{\tau_{2},s_{2,1}\}}\!<\! 1\right]
   \quad {\rm RTD} \\  
&\Pr\left[\frac{\gamma_1}{\min\{\tau_1,s_{1,1}\}}\!<\! 1,
    \frac{\gamma_1}{s_{1}}+
    \frac{\gamma_2}{\min\{\tau_{2},s_{2,1}\}}\Big(1\!+\!
    \theta  \frac{\gamma_1}{\tau_{1}}\Big)\!<\! 1\right] 
    \!\! {\rm INR}.  
\end{align*}
The probabilities for ALO and RTD can be evaluated in closed form, 
while the probabilities for INR can be bounded as in Subsection~\ref{ray M=2 F=2 old}
(from Proposition~\ref{prop:density of sumt gammat/taut} in Section~\ref{sect:main bounds})
for the classical HARQ protocols by using the function $q(x)$ in~\reff{eq:finally out pfd!}.

\subsubsection{Discussion}
Fig.~\ref{fig:thrratioszoom} shows the ratio among the throughput of new protocols and 
the ergodic water-filling capacity. We see that the new protocols
dramatically outperform the classical repetition protocols,
especially at low $\snr$ (compare with Fig.~\ref{fig:thrratio}).  Indeed, at low $\snr$ it is
critical to be able to save power when the channel is in deep fade.  By providing
the transmitter with a 1-bit quantization of the current channel gain (rather than ACK/NACK),
we enable the transmitter to do so.
%
At low $\snr$, the repetition is not needed (ALO with $M=1$ and $F=2$ has the
same throughput of INR with $M=2$ and $F=2$),
while at high $\snr$, the repetition helps. Notice that 
here high $\snr$ means is $\snr>5$dB, at which $\eta_{M=2,F=2}^{\rm(INR)}$
is about 67\% of $\eta_{M=\infty,F=\infty}^{\rm(INR)}$.
We did not report the RTD and ALO curve for $M=2$ and $F=2$
with CSI as they do not differ much from ALO with $M=1$ and $F=2$.

\section{Conclusions and Future Work}
\label{sect:conclusions}
In this work we considered HARQ protocols where
the feedback bits not only convey a retransmission
request to the transmitter but also inform the transmitter
coarsely about the channel state.  We developed a new 
class of protocols that feedback the quantized index of
a suitably scaled version of the current fading value;
the scaling factor is such that the mutual information already
accumulated at the transmitter from the previous transmissions
is taken into account.  
We showed that our proposed protocols significantly outperform 
classical HARQ protocols for the same amount of feedback
resources, especially at low $\snr$; this shows that ACK/NACK   
feedback is suboptimal in time-varying channels.

As future work, it would be interesting to evaluate the throughput performance
when the cost of acquiring the CSI and the error in the estimated CSI 
are taken into account.  Also, it is important to test the proposed
protocols with practical codes, instead of with ideal Gaussian codes.

Extensions to multiple access channels are presented in~\cite{tuninettiICC2011}.

\section*{Acknowledgment}
The author would like to thank the Associate Editor and the anonymous 
reviewers for their comments that helped to improve the quality of the
paper.

\appendices

\section{Proof of Theorem~\ref{thm:eta vs longtermpower}}
\label{sec:proof thm:eta vs longtermpower}
As a direct application of the renewal-reward theorem~\cite{tuninettiITW2007},
we have that the limit in~\reff{eq:power} converges almost surely to $\E[\Pc]/\E[\Tc]$
and the limit in~\reff{eq:throughput} converges almost surely to $\E[\Rc]/\E[\Tc]$.
The average inter-renewal time $\E[\Tc]$ is given by:
\begin{align*}
\E[\Tc]
  &= \sum_{m=1}^{M} m\,\Pr[\Tc=m]
   = \sum_{m=1}^{M} \Pr[\Tc\geq m].
\end{align*}
The average reward $\E[\Rc]$ is given by:
\begin{align*}
\E[\Rc]
  = R(1-P_{\rm out})
  = R(1- \Pr[\Tc= M, \, {\rm fail \, to \,decode}]).
\end{align*}
since in our framework, a packet is lost ($\Rc=0$) only if 
on the last transmission decoding was not successful.
The average transmit power $\E[\Pc]$ is given by:
\begin{align*}
\E[\Pc] 
  &=  \sum_{m=1}^{M} \Pr[\Tc=m] \E[\Pc|\,\Tc=m] 
\\&=  \sum_{m=1}^{M} \Pr[\Tc=m] \sum_{t=1}^{m}\E[P_t|\,\Tc=m] 
\\&=  \sum_{t=1}^{M} \sum_{m=t}^{M} \Pr[\Tc=m] \,\E[P_t|\,\Tc=m]
\\&=  \sum_{t=1}^{M} \Pr[\Tc\geq t] \frac{\sum_{m=t}^{M} \Pr[\Tc=m] \,\E[P_t|\,\Tc=m]}{\sum_{m=t}^{M} \Pr[\Tc=m] }
\\&=  \sum_{t=1}^{M} \Pr[\Tc\geq t]\,\E[P_t|\,\Tc\geq t],
\end{align*}
QED.
\section{Proof of Theorem~\ref{thm:Bhashyam intervals}}
\label{app:Bhashyam intervals}

Given a partition $\{\Rc_f\}$ of $\RR^+$,
the transmit power is defined through the
thresholds in~\reff{eq: def thresholds for eta M=infty,F,(INR)} as
\[
P = \sum_{f=1}^{F}\frac{\eu^{R}-1}{s_f}1_{\{\gamma\in\Rc_f\}},
\]
and must satisfy the average power constraint 
\[
\sum_{f=1}^{F}\frac{\eu^{R}-1}{s_f}\Pr[\gamma\in\Rc_f] \leq \overline{P}.
\]

For a fixed rate $R$, the throughput is maximized if the outage probability is minimized.
The outage probability satisfies
\begin{align*}
&\Pr\left[\log\left(1+\gamma\sum_{f=1}^{F}\frac{\eu^{R}-1}{s_f}1_{\{\gamma\in \Rc_f\}}\right)< R\right]
\\&=\E\left[1_{\{\log\left(1+\gamma\sum_{f=1}^{F}\frac{\eu^{R}-1}{s_f}1_{\{\gamma\in \Rc_f\}}\right)< R\}}\right]
\\&=\E\left[\sum_{f=1}^{F}1_{\{\gamma\in \Rc_f\}}
1_{\{ \gamma<s_f \}}\right]
\\&\geq \E\left[\min_{f=1,\ldots,F}1_{\{ \gamma<s_f \}}\right],
\end{align*}
where the last inequality holds with equality for
\[
\Rc^{\rm(ALO)}_f = \Big\{\gamma\in\RR^+: f=\arg\min_{\ell=1,\ldots,F}\big\{1_{\{ \gamma<s_\ell \}}\big\}\Big\}.
\]
By assuming without loss of generality that the thresholds $\{s_f\}$ are ordered in increasing order, we have:
\begin{align*}
&\arg\min_{f=1,\ldots,F}1_{\{ \gamma<s_f \}}\\
&\in\{1,\ldots,f\} \ {\rm if}\ \gamma\in[s_f,s_{f+1}), \ f=1,\ldots,F-1, \\
&\in\{1,\ldots,F\} \ {\rm if}\  \gamma\in[s_{F},+\infty)\cup[0,s_1).
\end{align*}
In order to use the least power we choose
\begin{align*}
&\arg\min_{f=1,\ldots,F}1_{\{ \gamma<s_f \}}\\
&=  f   \ {\rm if}\ \gamma\in R_f^{\rm(ALO)}=[s_f,s_{f+1}), f=1,\ldots.,F-1, \\ 
&=  F   \ {\rm if}\ \gamma\in R_F^{\rm(ALO)}=[s_{F},+\infty)\cup[0,s_1).
\end{align*}
With these quantization regions, an outage only occurs when 
the feedback value is $F$ and the fading is in $[0,s_1)$, hence
the throughput is
\begin{align*}
  &\eta_{M=1,F}^{\rm(ALO)}
\\&=\max_{R,\{s_f,\Rc_f\}} R
\Pr\left[R \leq \log\left(1+\gamma\sum_{f=1}^{F}\frac{\eu^{R}-1}{s_f}1_{\{\gamma\in \Rc_f\}}\right)\right]
\\&=\max_{\{s_f\}}
\log\left(1+\frac{\overline{P}}{\frac{\Pr[\gamma\in[0,s_1)]}{s_F}+\sum_{f=1}^{F}\frac{1}{s_f}\Pr[\gamma\in[s_f,s_{f+1})]}\right)
\cdot\\&\qquad \cdot
   \Pr[\gamma\geq s_1].
\end{align*}

\section{Proof of Proposition~\ref{prop: bounds eta M=1,F,(ALO)}}
\label{app:prop: bounds eta M=1,F,(ALO)}
The optimal values of $0\leq s_2\leq \ldots\leq s_{F}\leq s_{F+1}=\infty$ for
$\widehat{\eta}_{M=1,F}^{\rm (ALO)}$ in~\reff{eq: bound for eta M=1,F,(ALO)}
minimize 
\begin{align}
\sum_{f=1}^{F} \frac{1}{s_{f}} [F_\gamma(s_{f+1})-F_\gamma(s_{f})]
\label{eq:some power for alo m=1}.
\end{align}
By assuming that the fading has a density $f_\gamma(x)=\diff{F_\gamma(x)}/\diff{x}$,
by taking the partial derivatives of~\reff{eq:some power for alo m=1}
with respect to $s_\ell$ for $\ell\geq 2$, and solving them 
equal to zero, we get that the thresholds satisfy
\begin{align*}
\frac{s_{\ell} - s_{\ell-1}}{s_{\ell-1}}= \frac{F_\gamma(s_{\ell+1})-F_\gamma(s_\ell)}{s_{\ell}\,f_\gamma(s_{\ell})},
\quad \ell\geq 2.
\end{align*}
For sufficiently large $F$, since the thresholds are going to be close to each other,
we can approximate
\[
\frac{F_\gamma(s_{\ell+1})-F_\gamma(s_\ell)}{(s_{\ell+1}-s_{\ell})}
\approx
\left.\frac{\diff F_\gamma(x)}{\diff{x}}\right|_{x=s_\ell}
=
f_\gamma(s_{\ell}),
\]
and hence we conclude that the optimal thresholds satisfy 
\begin{align*}
s_\ell^2~\approx~s_{\ell-1}s_{\ell+1}
\Longleftrightarrow
s_\ell \approx s_1\,\xi^{\ell-1},
\end{align*}
for some  $\xi\geq1$.

\section{Proof of Proposition~\ref{prop:p  INR m bounds}}
\label{app:p  INR m bounds}
For a general $m\in\NN$, the probability in~\reff{eq:pminr}
(with $X_\ell=\gamma_\ell/\tau_\ell$)
is equivalent to
\[
p_m^{\rm (INR)}=\Pr[X_m - f(X_1,\ldots,X_{m-1})<0],
\]
for
\[
f(x_1,\ldots,x_{m-1})=
\frac{1}{\theta}\left(\frac{1+\theta}{
\prod_{s=1}^{m-1}\left(1+\theta x_s\right)
} - 1 \right).
\]
Our goal is to bound the region below $f(x_1,\ldots,x_{m-1})$ in the positive
quadrant by regions defined as union of hyperplanes.   Toward this goal,
it is useful to keep in mind that the partial derivatives of $f(x_1,\ldots,x_{m-1})$
are given by
\[
\frac{\partial f(x_1,\ldots,x_{m-1})}{\partial x_j}
= - \frac{1+\theta\, f(x_1,\ldots,x_{m-1})}{1+\theta x_j}.
\]
Being $f(x_1,\ldots,x_{m-1})$ a concave function in all its arguments,
it can be lower-bounded by any hyperplane tangent to it.
In particular, if we consider the hyperplanes tangent to $f(x_1,\ldots,x_{m-1})$
at those points with at most one non-zero coordinate at value~1, we obtain:
\[
f(x_1,\ldots,x_{m-1}) \geq
- \sum_{s=1, \ldots, m-1, \, s\not= t} \left(x_s - 0\right)
- \frac{1}{1+\theta} \left(x_t - 1\right),
\]
for all $t=1,\ldots,m$,
which is equivalent to
\[
x_m - f(x_1,\ldots,x_{m-1}) \leq
\sum_{s=1}^{m} x_s - \frac{1+ x_t\,\theta}{1+\theta}, \quad \forall t=1,\ldots,m.
\]
This bound implies
\begin{align*}
&p_m^{\rm (INR)}
  \geq \Pr\left[ \bigcup_{t=1}^{m} \,\,
 (1+\theta) \sum_{s=1}^{m}\frac{\gamma_s}{\tau_s}
 < 1+\theta \frac{\gamma_t}{\tau_t}
  \right]
\nonumber
\\&=  \Pr\left[ 
 (1+\theta) \sum_{s=1}^{m}\frac{\gamma_s}{\tau_s}
 < 1+\theta \, \max_{t=1,\ldots,m} \frac{\gamma_t}{\tau_t}
  \right]
\\&= {\rm\reff{eq:p  INR m geq}}.
\end{align*}

In the same spirit, we bound $f(x_1,\ldots,x_{m-1})$ from above
by considering the union of the region
below the $m$ hyper-planes defined as follows:  for a fixed $t\in\{1,\ldots,m\}$, 
the hyper-plane that passes through the points with coordinates
$x_k=1$ and $x_j=0$ for all $j\not= k$ and $k\not= t$
(this defines $m-1$ points with only a single non-zero coordinate at 1)
and the point $x_1= \ldots = x_m = \frac{(1+\theta)^{1/m}-1}{\theta}$ is
defined as
\[
\sum_{s=1}^{m} x_s + \left(\frac{\theta}{(1+\theta)^{1/m}-1}-m\right) x_t = 1.
\]
The region below the union of the above hyperplanes for all $t=1,\ldots,m$ contains the
region that defines $p_m^{\rm (INR)}$ and hence
\begin{align*}
&p_m^{\rm (INR)}
  \leq \Pr\left[   \bigcup_{t=1}^{m} \,\,
 \sum_{s=1}^{m}\frac{\gamma_s}{\tau_s}
 + \left(\frac{\theta}{(1+\theta)^{1/m}-1}-m\right) \frac{\gamma_t}{\tau_t} < 1
  \right]
\nonumber
\\&=
\Pr\left[ 
 \sum_{s=1}^{m}\frac{\gamma_s}{\tau_s}
 + \left(\frac{\theta}{(1+\theta)^{1/m}-1}-m\right) \max_{t=1,\ldots,m} \frac{\gamma_t}{\tau_t}
 < 1
  \right]
  \\&= {\rm\reff{eq:p  INR m leq}}.
\end{align*}

\section{Proof of Proposition~\ref{prop:density of sumt gammat/taut}}
\label{app:orderstst}
It is a well known result in order statistics~\cite[Ch.5]{book:davidorderstat}
that the order statistics of a sample of independent negative exponential random variables can be expressed as the unorder statistics of a sample of independent negative exponential random variables with appropriate mean value. We report the derivation of result here for sake of completeness. 

Let $\{\gamma_k\}_{k=1}^{K}$ be an independent sample of size $K$ of negative exponential distributed random variables with mean $\E[\gamma_k]=\mu_k$.
Let $\pi$ be a permutation of the integers $\{1,2,\ldots,K\}$ and let $\Pc_K$ be the set of the $K!$ such permutations. With an abuse of notation, we also indicate with $\pi$ the event
\[
\Pr[\pi] \defeq
\Pr[\gamma_{\pi(1)}\geq \gamma_{\pi(2)}\geq \ldots \geq \gamma_{\pi(K)}],
\quad \forall \pi\in\Pc_K.
\]
It is well known that
\begin{align*}
&f_{\gamma_1,\ldots,\gamma_K|\pi}(x_1,\ldots,x_K)
\\&= \frac{1}{\Pr[\pi]} \prod_{i=1}^{K}
  \frac{1}{\mu_i} \exp\Big(-\frac{x_{i}}{\mu_i}\Big)
 \, 1_{ \{ x_{\pi(1)}\geq\ldots\geq x_{\pi(K)}\geq0 \} }.
\end{align*}
Fix a permutation $\pi$, define $\gamma_{\pi(K+1)}=0$, and
consider the following change of variables with $c^{(\pi)}_i >0$ for all $i=1,\ldots,K$:
\begin{align*}
Z_i \defeq c^{(\pi)}_i(\gamma_{\pi(i)} - \gamma_{\pi(i+1)})
\Longleftrightarrow 
\gamma_{\pi(i)} = \sum_{j=i}^{K}\frac{Z_j}{c^{(\pi)}_j}.
\end{align*}
The random variables $\{Z_k\}_{k=1}^{K}$ are non-negative by definition.
Moreover, the transformation giving $\{Z_k\}_{k=1}^{K}$ from $\{\gamma_k\}_{k=1}^{K}$ has Jacobian $\prod_{i=1}^{K}c^{(\pi)}_i$.  Hence, the joint density of $\{Z_k\}_{k=1}^{K}$ is
\begin{align*}
  &f_{Z_{1},\ldots,Z_{K}| \pi}(z_1,\ldots,z_n)
\\&= \frac{1}{\Pr[\pi]} 
  \frac{1}{\prod_{i=1}^{K} \mu_i\,\, c^{(\pi)}_i}
  \exp\left(-\sum_{k=1}^{K} \frac{1}{\mu_{\pi(k)}}
  \sum_{j=k}^{K}   \frac{z_{j}}{c^{(\pi)}_{j}}\right) 
  \,\prod_{i=1}^{K}\,1_{\{z_i\geq0\}}
\\&=
  \prod_{j=1}^{K} \frac{1}{\theta^{(\pi)}_{j}}
  \exp\Big(-\frac{z_j}{\theta^{(\pi)}_{j}}\Big) 1_{\{z_j\geq0\}},
\end{align*}
where 
\begin{align*}
   \theta^{(\pi)}_{j} &\defeq
   \frac{c^{(\pi)}_j}{\sum_{k=1}^{j}\frac{1}{\mu_{\pi(k)}}},
\end{align*}
obtained by recalling that
\[
\frac{1}{\Pr[\pi]} =
    \prod_{j=1}^{K} \left({\sum_{k=1}^{j}\frac{\mu_{\pi(j)}}{\mu_{\pi(k)}}}\right).
\]
In words, the change of variables has produced the unorder statistics of an independent sample of size $K$ of negative exponential distributed random variables with mean $\E[Z_k]=\theta^{(\pi)}_k$.  Remark: in the case $\mu_i=\mu$ for all $i=1,\ldots,K$
one obtains the familiar result (think at the inter-arrival times of a Poisson process):
\begin{align*}
 \frac{\theta^{(\pi)}_{j}}{c^{(\pi)}_j} = \frac{\mu}{j},
\quad
 \frac{1}{\Pr[\pi]} =  K!
\end{align*}

\medskip
We now apply this result to the computation of the distribution of
\[
X_{a,b} \defeq a\,\max_{k=1,\ldots,K} \gamma_k 
        + b\,\sum_{k=1}^{K} \gamma_k
\]
For general $a$ and $b$ we have: 
\begin{align*}
  &\Pr[X_{a,b} \leq x]
\\&= \Pr\left[ a\,\max_{k=1,\ldots,K} \gamma_k 
        + b\,\sum_{k=1}^{K} \gamma_k \leq x \right]
\\&= \sum_{\pi\in\Pc_K}\Pr[\pi]
     \Pr\left[ a\,\gamma_{\pi(1)}
              + b\,\sum_{k=1}^{K} \gamma_{\pi(k)} 
              \leq x\,|\,\pi\right]
\\&= \sum_{\pi\in\Pc_K}\Pr[\pi]
     \Pr\left[ a\,\sum_{\ell=1}^{K}\frac{Z^{(\pi)}_\ell}{c^{(\pi)}_\ell}
              + b\,\sum_{k=1}^{K} \sum_{\ell=k}^{K}\frac{Z^{(\pi)}_\ell}{c^{(\pi)}_\ell} 
              \leq x\,|\,\pi\right]
\\&= \sum_{\pi\in\Pc_K}\Pr[\pi]
     \Pr\left[ a\,\sum_{\ell=1}^{K}\frac{Z^{(\pi)}_\ell}{c^{(\pi)}_\ell}           
              + b\,\sum_{\ell=1}^{K} \ell\,\frac{Z^{(\pi)}_\ell}{c^{(\pi)}_\ell} 
              \leq x\,|\,\pi\right].
\end{align*}
We next chose $c^{(\pi)}_\ell = \sum_{k=1}^{\ell}\frac{1}{\mu_{\pi(k)}}$
so that the $Z^{(\pi)}_{\ell}$ are iid negative exponential with unit mean value
for all $\ell\in\{1,\ldots,K\}$ and for all $\pi\in\Pc_K$, hence we obtain
\[
\Pr[X_{a,b} \leq x]
= \sum_{\pi\in\Pc_K}\Pr[\pi]
     \Pr\left[ \sum_{\ell=1}^{K}Z_\ell\frac{a+b\,\ell}{\sum_{k=1}^{\ell}\frac{1}{\mu_{\pi(k)}}}
     \leq x \right]
\]
At this point, the problem reduces to that of finding the density of a random variable $Y$ that is a linear combination of iid negative exponential with unit mean random variables, i.e.,  $Y = \sum_{k=1}^{K} v_k Z_k$ for $v_k\in\RR$ for all $k=1,\ldots,K$ .
The calculus of residues applied to the characteristic function of $Y$,
assuming that all the coefficients $v_i$ are distinct, gives:
\begin{align*}
  g_Y(\omega)
  &\defeq \E[\exp(-\jj \omega \sum_{k=1}^{K} v_k Z_k) ] 
\\&= \prod_{k=1}^{K} \E[\exp(-\jj \omega v_k Z_k) ]
\\&= \prod_{k=1}^{K} \frac{1}{1+\jj \omega v_k }
   = \sum_{k=1}^{K} \frac{\alpha_k}{1+\jj \omega v_k},
    \\&\text{with}
   \quad \alpha_k \defeq {\rm Residue}[g_Y(\omega),v_k]
     = \prod_{\ell\not=k, \ell=1}^{K}\frac{1}{1-v_\ell/v_k},
    \\&\text{such that}
      \quad \sum_{k=1}^{K} \alpha_k =1.
\end{align*}
By taking the inverse Fourier transform of $g_Y(\omega)$
we obtain the density function of $Y$ given by
\begin{align*}
f_Y(x)  
  =\sum_{k=1}^{K}
  \frac{\alpha_k}{|v_k|}
  \eu^{-x/v_k}\, 1_{\{x\,{\rm sign}(v_k) \geq 0\}}
\end{align*}
and hence
\begin{align*}
  \Pr\big[\sum_{k=1}^{K} v_k Z_k \leq x\big]
  &=\Big[1-\sum_{k=1: v_k> 0}^{K} \alpha_k  \eu^{-x/v_k}\Big]\, 1_{\{x \geq 0\}}+
\\&+\Big[  \sum_{k=1: v_k< 0}^{K} \alpha_k  \eu^{-x/v_k}\Big]\, 1_{\{x \leq 0\}}
\end{align*}
If some coefficients are equal, say $v_1$ and $v_2$, then it suffices to late the limit for $v_1\to v_2$ of $\Pr\big[\sum_{k=1}^{K} v_k Z_k \leq x\big]$.

\medskip
Back to our original problem:
\begin{align*}
\Pr[X_{a,b} \leq x]
  &=\sum_{\pi\in\Pc_K}\Pr[\pi] \Big[
  \Big(1-\sum_{\ell: v^{(\pi)}_\ell >0}
         \alpha^{(\pi)}_\ell \eu^{-x/v^{(\pi)}_\ell} \Big) 1_{\{x \geq 0\}}+
\\&+\Big(\sum_{\ell: v^{(\pi)}_\ell < 0}
         \alpha^{(\pi)}_\ell \eu^{-x/v^{(\pi)}_\ell} \Big) 1_{\{x \leq 0\}}
 \Big],
\\v^{(\pi)}_\ell &\defeq \frac{a+b\,\ell}{\sum_{k=1}^{\ell}\frac{1}{\mu_{\pi(k)}}},
\\\alpha^{(\pi)}_\ell &\defeq \prod_{j\not = \ell, j=1}^{K} \frac{1}{1-v^{(\pi)}_j/v^{(\pi)}_\ell}
\end{align*}
%

For example, with $b>0$ and $a+b>0$, so that $a+b \ell>0$ for all $\ell\in\NN$
(this is the case of interest in our problem),
for $K=1$: $X_{a,b}= (a+b)\gamma_1$, for $x\geq 0$
 \begin{align*}
1 -F_{X_{a,b}} (x)  
 = \eu^{-x\frac{1}{(a+b)\mu_1}}
\end{align*}
For $K=2$: $X_{a,b} = (a+b)\max\{\gamma_1,\gamma_2\} + b \min\{\gamma_1,\gamma_2\}$,
for $x\geq 0$
\begin{align}
&1 - F_{X_{a,b}} (x)  
\nonumber\\&=
    \eu^{-x\tau_1} \frac{c\,\tau_2}{c\,\tau_1 + c\,\tau_2 - \tau_1}
   +\eu^{-x\tau_2} \frac{c\,\tau_1}{c\,\tau_1 + c\,\tau_2 - \tau_2}
   +
\nonumber\\&
   +\eu^{-c\,x(\tau_1+\tau_2)} \frac{\tau_1 \tau_2(1-2c)}
{(c\,\tau_1 + c\,\tau_2 - \tau_1)(c\,\tau_1 + c\,\tau_2 - \tau_2)},
\nonumber\\&
{\tau_k = \frac{1}{(a+b)\mu_k},\, c=\frac{a+b}{a+2b}}.
\label{eq:finally out pfd!}
\end{align}


\begin{thebibliography}{10}
\providecommand{\url}[1]{#1}
\csname url@samestyle\endcsname
\providecommand{\newblock}{\relax}
\providecommand{\bibinfo}[2]{#2}
\providecommand{\BIBentrySTDinterwordspacing}{\spaceskip=0pt\relax}
\providecommand{\BIBentryALTinterwordstretchfactor}{4}
\providecommand{\BIBentryALTinterwordspacing}{\spaceskip=\fontdimen2\font plus
\BIBentryALTinterwordstretchfactor\fontdimen3\font minus
  \fontdimen4\font\relax}
\providecommand{\BIBforeignlanguage}[2]{{%
\expandafter\ifx\csname l@#1\endcsname\relax
\typeout{** WARNING: IEEEtran.bst: No hyphenation pattern has been}%
\typeout{** loaded for the language `#1'. Using the pattern for}%
\typeout{** the default language instead.}%
\else
\language=\csname l@#1\endcsname
\fi
#2}}
\providecommand{\BIBdecl}{\relax}
\BIBdecl

\bibitem{tuninettiICC2008}
J.~Perret and D.~Tuninetti, ``Repetition protocols for block fading channels
  that combine transmission requests and state information,'' in
  \emph{Proceedings of ICC Int. Conf. Comm., ICC2008}, Beijing, China, May
  2008.

\bibitem{tuninettiITW2007}
D.~Tuninetti, ``Transmitter channel state information and repetition protocols
  in block fading channels,'' in \emph{Proceedings of IEEE Int. Workshop. on
  Inform. Theory, ITW2007}, Lake Tahoe, CA, September 2007.

\bibitem{caire_tuninetti:arq_it}
G.~Caire and D.~Tuninetti, ``The throughput of {H}ybrid-{ARQ} protocols for the
  {G}aussian collision channel,'' \emph{IEEE Trans.\ Inform.\ Theory}, vol.~47,
  no.~5, pp. 1971--1988, July 2001.

\bibitem{goldsmith_varaiya}
A.~Goldsmith and P.~Varaiya, ``Capacity of fading channels with channel state
  information,'' \emph{IEEE Trans.\ Inform.\ Theory}, vol.~43, no.~6, pp.
  1986--1992, November 1997.

\bibitem{summa_fading_bc}
E.Biglieri, J.Proakis, and S.Shamai, ``Fading channels: information-theoretic
  and communications aspects,'' \emph{IEEE Trans.\ Inform.\ Theory}, vol.~44,
  no.~6, pp. 2619 --2692, Oct. 1998.

\bibitem{Kim-Skolunt}
T.~T. Kim and M.~Skoglund, ``On the expected rate of slowly fading channels
  with quantized side information,'' \emph{IEEE Trans. on Commun.}, vol.~55,
  no.~4, pp. 820 -- 829, April 2007.

\bibitem{elgamalcairedamen:dmtharq:it06}
H.~E. Gamal, G.~Caire, and M.~O. Damen, ``The mimo arq channel:
  Diversity-multiplexing-delay tradeoff,'' \emph{IEEE Trans.\ Inform.\ Theory},
  vol.~52, no.~8, pp. 3601 -- 3621, August 2006.

\bibitem{Bhashyam}
S.~Bhashyam, A.~Sabharwal, and B.~Aazhang, ``Feedback gain in multiple antenna
  systems,'' \emph{IEEE Trans. on Commun.}, vol.~50, no.~5, pp. 785 -- 798, May
  2002.

\bibitem{steinber-samai}
A.~Steiner and S.~S. (Shitz), ``Broadcasting with partial transmit channel
  state information,'' \emph{Joint NEWCOM-ARoC Workshop}, Sep. 2006.

\bibitem{caireshamai:capacitypartialcsi}
G.~Caire and S.~S. (Shitz), ``On the capacity of some channels with channel
  state information,'' \emph{IEEE Trans.\ Inform.\ Theory}, vol.~45, no.~6, pp.
  2007 -- 2019, 1999.

\bibitem{caire_tuninetti_verdu:rate_it}
G.~Caire, D.~Tuninetti, and S.~Verd\'u, ``Variable-rate coding for
  slowly-fading gaussian channels,'' \emph{IEEE Trans.\ Inform.\ Theory},
  vol.~50, no.~10, pp. 2271--2292, October 2004.

\bibitem{caire_taricco_biglieri:optimal_powercontrol}
G.~Caire, G.~Taricco, and E.~Biglieri, ``Optimum power control over fading
  channel,'' \emph{IEEE Trans.\ Inform.\ Theory}, vol.~45, no.~5, pp.
  1468--1489, July 1999.

\bibitem{book:kumar_varaiya:stochastic_systems}
R.~Kumar and P.~Varaiya, \emph{Stochastic Systems: estimation, identification
  and adaptive control}.\hskip 1em plus 0.5em minus 0.4em\relax Prentice-Hall,
  1986.

\bibitem{book:goldsmith:wireless}
A.~Goldsmith, \emph{Wireless Communications}.\hskip 1em plus 0.5em minus
  0.4em\relax Cambridge University Press, 2005.

\bibitem{negi_cioffi:dymprog_jrnl}
R.Negi and J.Cioffi, ``Delay-constrained capacity with causal feedback,''
  \emph{IEEE Trans.\ Inform.\ Theory}, vol.~48, no.~9, pp. 2478 --2494,
  September 2002.

\bibitem{book:grimmet_strizaker:prob}
G.R.Grimmet and D.R.Strizaker, \emph{Probability and Random Processes},
  2nd~ed.\hskip 1em plus 0.5em minus 0.4em\relax New York: Oxford University
  Press, 1992.

\bibitem{lloydalgorithm-57}
S.~Lloyd, ``Least squares quantization in pcm (reprint of unpublished bell lab.
  note, sept. 1957),'' \emph{IEEE Trans.\ Inform.\ Theory}, vol.~28, pp.
  127--135, March 1982.

\bibitem{tuninettiICC2011}
D.~Barbieri and D.~Tuninetti, ``On repetition protocols and power control for
  multiple access block-fading channels,'' in \emph{Proceedings of ICC Int.
  Conf. Comm., ICC2011}, Kyoto, Japan, June 2011.

\bibitem{book:davidorderstat}
H.~David and H.~N. Nagaraja, \emph{Ordered Statistics, 3rd edition}.\hskip 1em
  plus 0.5em minus 0.4em\relax New York: Wiley, 199?

\end{thebibliography}

\end{document}